\input harvmac
\noblackbox
\input epsf
\newcount\figno
\figno=0
\def\fig#1#2#3{
\par\begingroup\parindent=0pt\leftskip=1cm\rightskip=1cm\parindent=0pt
\baselineskip=11pt \global\advance\figno by 1 \midinsert
\epsfxsize=#3 \centerline{\epsfbox{#2}} \vskip 12pt
\centerline{{\bf Figure \the\figno:} #1}\par
\endinsert\endgroup\par}
\def\figlabel#1{\xdef#1{\the\figno}}

\def\np#1#2#3{Nucl. Phys. {\bf B#1} (#2) #3}

\def\IR{\relax{\rm I\kern-.18em R}}


\font\cmss=cmss10 \font\cmsss=cmss10 at 7pt
\def\rlx{\relax\leavevmode}
\def\inbar{\vrule height1.5ex width.4pt depth0pt}
\def\IC{\relax\,\hbox{$\inbar\kern-.3em{\rm C}$}}
\def\IN{\relax{\rm I\kern-.18em N}}
\def\IP{\relax{\rm I\kern-.18em P}}
\def\ZZ{\rlx\leavevmode\ifmmode\mathchoice{\hbox{\cmss Z\kern-.4em Z}}
 {\hbox{\cmss Z\kern-.4em Z}}{\lower.9pt\hbox{\cmsss Z\kern-.36em Z}}
 {\lower1.2pt\hbox{\cmsss Z\kern-.36em Z}}\else{\cmss Z\kern-.4em
 Z}\fi}
\def\IZ{\relax\ifmmode\mathchoice
{\hbox{\cmss Z\kern-.4em Z}}{\hbox{\cmss Z\kern-.4em Z}}
{\lower.9pt\hbox{\cmsss Z\kern-.4em Z}} {\lower1.2pt\hbox{\cmsss
Z\kern-.4em Z}}\else{\cmss Z\kern-.4em Z}\fi}

\def\narrowplus{\kern -.04truein + \kern -.03truein}
\def\narrowminus{- \kern -.04truein}
\def\narrowminussub{\kern -.02truein - \kern -.01truein}

\def\a{{\alpha}}

\def\frac#1#2{{#1\over #2}}

\def\IZ{\relax\ifmmode\mathchoice
{\hbox{\cmss Z\kern-.4em Z}}{\hbox{\cmss Z\kern-.4em Z}}
{\lower.9pt\hbox{\cmsss Z\kern-.4em Z}} {\lower1.2pt\hbox{\cmsss
Z\kern-.4em Z}}\else{\cmss Z\kern-.4em Z}\fi}
\def\IB{\relax{\rm I\kern-.18em B}}
\def\IC{{\relax\hbox{$\inbar\kern-.3em{\rm C}$}}}
\def\ID{\relax{\rm I\kern-.18em D}}
\def\IE{\relax{\rm I\kern-.18em E}}
\def\IF{\relax{\rm I\kern-.18em F}}
\def\IG{\relax\hbox{$\inbar\kern-.3em{\rm G}$}}
\def\IGa{\relax\hbox{${\rm I}\kern-.18em\Gamma$}}
\def\IH{\relax{\rm I\kern-.18em H}}
\def\II{\relax{\rm I\kern-.18em I}}
\def\IK{\relax{\rm I\kern-.18em K}}
\def\IP{\relax{\rm I\kern-.18em P}}

\font\cmss=cmss10 \font\cmsss=cmss10 at 7pt
\def\IR{\relax{\rm I\kern-.18em R}}

\def\1{{\bf 1}}
\def\3{{\bf 3}}
\def\7{{\bf 7}}
\def\2{{\bf 2}}
\def\8{{\bf 8}}

\def\hat{\widehat}

\def\o{\over}
%

%
%
\def\eqnn#1{\xdef #1{(\secsym\the\meqno)}\writedef{#1\leftbracket#1}%
\global\advance\meqno by1\wrlabeL#1}
\def\eqna#1{\xdef #1##1{\hbox{$(\secsym\the\meqno##1)$}}
\writedef{#1\numbersign1\leftbracket#1{\numbersign1}}%
\global\advance\meqno by1\wrlabeL{#1$\{\}$}}
\def\eqn#1#2{\xdef #1{(\secsym\the\meqno)}\writedef{#1\leftbracket#1}%
\global\advance\meqno by1$$#2\eqno#1\eqlabeL#1$$}


\lref\DuffWD{ M.~J.~Duff, J.~T.~Liu and R.~Minasian, {\it
``Eleven-Dimensional Origin of String/String Duality: A One-Loop
Test''}, Nucl.\ Phys. {\bf B452} (1995) 261, hep-th/9506126. }

\lref\rBB{ K.~Becker and M.~Becker, {\it ``${\cal M}$-Theory on
Eight-Manifolds,''}, Nucl.\ Phys.\ {\bf B477} (1996) 155,
hep-th/9605053.}

\lref\DasguptaSS{ K.~Dasgupta, G.~Rajesh and S.~Sethi, {\it ``M
theory, Orientifolds and $G$-flux''}, JHEP {\bf 9908} (1999) 023,
hep-th/9908088. }

\lref\beckerD{ K.~Becker and K.~Dasgupta, {\it ``Heterotic
Strings with Torsion,''} hep-th/0209077.}

\lref\banks{T. Banks, ``Cosmological Breaking of Supersymmetry?
or little Lambda Goes Back to the Future'', hep-th/0007146.}

\lref\rBHO{E. Bergshoeff, C. Hull and T. Ortin, {\it ``Duality in
the Type II Superstring Effective Action''}, \np{451}
{1995}{547}, hep-th/9504081. }
 \lref\rsenorien{A. Sen, {\it ``F-theory and Orientifolds''},
 Nucl. Phys. {\bf B475} (1996) 562,
hep-th/9605150.}

\lref\rstrom{A. ~Strominger, {\it ``Superstrings With Torsion''},
Nucl.\ Phys.\ {\bf B274} (1986) 253.}

\lref\witteno{E.~Witten, {\it ``Some Properties Of O(32)
Superstrings,''} Phys.\ Lett.\ B {\bf 149}, 351 (1984).}

\lref\sencount{A.~Sen, {\it ``Local Gauge and Lorentz Invariance
of the Heterotic String Theory,''} Phys.\ Lett.\ B {\bf 166}, 300
(1986)}

\lref\xenwit{M.~Dine, N.~Seiberg, X.~G.~Wen and E.~Witten, {\it
``Nonperturbative Effects on the String World Sheet,''} Nucl.\
Phys.\ B {\bf 278}, 769 (1986); {\it ``Nonperturbative Effects on
the String World Sheet. 2,''} Nucl.\ Phys.\ B {\bf 289}, 319
(1987).}

\lref\louisL{S.~Gurrieri, J.~Louis, A.~Micu and D.~Waldram, {\it
``Mirror Symmetry in Generalized Calabi-Yau Compactifications,''}
hep-th/0211102}

\lref\HULL{C.~M.~Hull, {\it ``Superstring Compactifications with
Torsion and Space-Time Supersymmetry,''} In Turin 1985,
Proceedings, Superunification and Extra Dimensions, 347-375, 29p;
{\it ``Sigma Model Beta Functions and String Compactifications,''}
Nucl.\ Phys.\ B {\bf 267}, 266 (1986); {\it ``Compactifications of
the Heterotic Superstring,''} Phys.\ Lett.\ B {\bf 178}, 357
(1986); {\it ``Lectures on Nonlinear Sigma Models and Strings,''}
Lectures given at Super Field Theories Workshop, Vancouver,
Canada, Jul 25 - Aug 6, 1986.}

\lref\hetcit{D.~J.~Gross, J.~A.~Harvey, E.~J.~Martinec and
R.~Rohm, {\it ``The Heterotic String,''} Phys.\ Rev.\ Lett.\
{\bf 54}, 502 (1985); {\it ``Heterotic String Theory. 1. The Free
Heterotic String,''} Nucl.\ Phys.\ B {\bf 256}, 253 (1985); {\it
``Heterotic String Theory. 2. The Interacting Heterotic String,''}
Nucl.\ Phys.\ B {\bf 267}, 75 (1986).}

\lref\rDJM{A.~Sen, {\it ``Strong Coupling Dynamics of Branes from
M-theory,''} JHEP {\bf 9710}, 002 (1997), 9708002; K. Dasgupta,
D. P. Jatkar and S. Mukhi, {\it ``Gravitational Couplings and
$Z_2$ Orientifolds''}, Nucl. Phys. {\bf B523} (1998) 465,
hep-th/9707224; J.~F.~Morales, C.~A.~Scrucca and M.~Serone, {\it
``Anomalous Couplings for D-branes and O-planes,''} Nucl.\ Phys.\
B {\bf 552}, 291 (1999), hep-th/9812071; B.~J.~Stefanski, {\it
``Gravitational Couplings of D-branes and O-planes,''} Nucl.\
Phys.\ B {\bf 548}, 275 (1999), hep-th/9812088. }

\lref\nemanja{N.~Kaloper and R.~C.~Myers, {\it ``The O(dd) Story
of Massive Supergravity,''} JHEP {\bf 9905}, 010 (1999),
hep-th/9901045; G.~Curio, A.~Klemm, B.~Kors and D.~Lust, {\it
``Fluxes in Heterotic and Type II String Compactifications,''}
Nucl.\ Phys.\ B {\bf 620}, 237 (2002), hep-th/0106155; J.~Louis
and A.~Micu, {\it ``Heterotic String Theory with Background
Fluxes,''} Nucl.\ Phys.\ B {\bf 626}, 26 (2002), hep-th/0110187.}

\lref\kapulov{V.~Kaplunovsky, J.~Louis and S.~Theisen, {\it
``Aspects of Duality in N=2 String Vacua,''} Phys.\ Lett.\ B {\bf
357}, 71 (1995); hep-th/9506110.}

\lref\nati{N.~Seiberg, {\it ``Observations on the Moduli Space of
Superconformal Field Theories,''} Nucl.\ Phys.\ B {\bf 303}, 286
(1988); A.~Ceresole, R.~D'Auria, S.~Ferrara and A.~Van Proeyen,
{\it ``Duality Transformations in Supersymmetric Yang-Mills
Theories Coupled to Supergravity,''} Nucl.\ Phys.\ B {\bf 444}, 92
(1995), hep-th/9502072.}

\lref\narainJ{K.~S.~Narain, {\it ``New Heterotic String Theories
in Uncompactified Dimensions $< 10$,''} Phys.\ Lett.\ B {\bf 169},
41 (1986); K.~S.~Narain, M.~H.~Sarmadi and E.~Witten, {\it ``A
Note on Toroidal Compactification of Heterotic String Theory,''}
Nucl.\ Phys.\ B {\bf 279}, 369 (1987); J.~Maharana and
J.~H.~Schwarz, {\it ``Noncompact Symmetries in String Theory,''}
Nucl.\ Phys.\ B {\bf 390}, 3 (1993), hep-th/9207016.}

\lref\senM{A.~Sen, {\it ``A Note on Enhanced Gauge Symmetries in
M- and String Theory,''} JHEP {\bf 9709}, 001 (1997),
hep-th/9707123}

\lref\rBUSH{T. Buscher, {\it ``Quantum Corrections and Extended
Supersymmetry in New Sigma Models''}, Phys. Lett. {\bf B159}
(1985) 127; {\it ``A Symmetry of the String Background Field
Equations''}, Phys. Lett. {\bf B194} (1987) 59; {\it ``Path
Integral Derivation of Quantum Duality in Nonlinear Sigma
Models''}, Phys. Lett. {\bf B201} (1988) 466.}

\lref\rKKL{E. Kiritsis, C. Kounnas and D. Lust, {\it ``A Large
Class of New Gravitational and Axionic Backgrounds for
Four-dimensional Superstrings''}, Int. J. Mod. Phys. {\bf A9}
(1994) 1361, hep-th/9308124. }

\lref\gauntlett{J.~P.~Gauntlett, N.~w.~Kim, D.~Martelli and D.~Waldram,
{\it ``Fivebranes wrapped on SLAG three-cycles and related geometry,''}
JHEP {\bf 0111}, 018 (2001) hep-th/0110034.
J.~P.~Gauntlett, D.~Martelli, S.~Pakis and D.~Waldram,
{\it ``G-structures and wrapped NS5-branes,''} hep-th/0205050;
J.~P.~Gauntlett, D.~Martelli and D.~Waldram,
{\it ``Superstrings with intrinsic torsion,''} hep-th/0302158.}

\lref\kst{S. Giddings, S. Kachru and J. Polchinski, {\it
``Hierarchies {}From Fluxes in String Compactifications''},
hep-th/0105097; S.~Kachru, M.~B.~Schulz and S.~Trivedi, {\it
``Moduli Stabilization from Fluxes in a Simple IIB Orientifold''},
hep-th/0201028; A.~R.~Frey and J.~Polchinski, {\it ``N = 3 Warped
Compactifications''}, Phys.\ Rev.\ {\bf D65} (2002) 126009,
hep-th/0201029. }

\lref\kstt{S.~Kachru, M.~B.~Schulz, P.~K.~Tripathy and
S.~P.~Trivedi, {\it ``New Supersymmetric String
Compactifications,''} hep-th/0211182.}

\lref\pktspto{ S.~Gurrieri and A.~Micu, {\it ``Type IIB theory on
half-flat manifolds,''} hep-th/0212278.}
\lref\pktsptt{P.~K.~Tripathy and S.~P.~Trivedi, {\it
Compactifications with Flux on K3 and Tori,''} hep-th/0301139.}

\lref\gates{ S.~J.~Gates, {\it ``Superspace Formulation Of New
Nonlinear Sigma Models,''} Nucl.\ Phys.\ B {\bf 238}, 349 (1984);
S.~J.~Gates, C.~M.~Hull and M.~Rocek, {\it ``Twisted Multiplets
and New Supersymmetric Nonlinear Sigma Models,''} Nucl.\ Phys.\ B
{\bf 248}, 157 (1984); S.~J.~Gates, S.~Gukov and E.~Witten, {\it
``Two-dimensional Supergravity Theories from Calabi-Yau
Four-folds,''} Nucl.\ Phys.\ B {\bf 584}, 109 (2000),
hep-th/0005120.}

\lref\rkehagias{A. Kehagias, {\it ``New Type IIB Vacua and their
F-theory Interpretation''}, Phys. Lett. {\bf B435} (1998) 337,
hep-th/9805131. }

\lref\GukovYA{ S.~Gukov, C.~Vafa and E.~Witten, {\it ``CFT's from
Calabi-Yau Four-folds''}, Nucl.\ Phys.\ {\bf B584} (2000) 69,
hep-th/9906070. }
\lref\BeckerPM{ K.~Becker and M.~Becker, {\it ``Supersymmetry
Breaking, ${\cal M}$-theory and Fluxes''}, JHEP {\bf 0107} (2001)
038 (2001), hep-th/0107044. }
\lref\DineRZ{ M.~Dine, R.~Rohm, N.~Seiberg and E.~Witten, {\it
``Gluino Condensation in Superstring Models''}, Phys.\ Lett.\ {\bf
B156}, 55 (1985).}
\lref\KachruHE{ S.~Kachru, M.~B.~Schulz and S.~Trivedi, {\it
``Moduli Stabilization from Fluxes in a Simple IIB Orientifold''},
hep-th/0201028.}
\lref\FreyHF{ A.~R.~Frey and J.~Polchinski, {\it ``N = 3 Warped
Compactifications''}, Phys.\ Rev.\ {\bf D65} (2002) 126009,
hep-th/0201029.}
\lref\CurioAE{ G.~Curio, A.~Klemm, B.~Kors and D.~Lust, {\it
``Fluxes in Heterotic and Type II String Compactifications''},
Nucl.\ Phys.\ {\bf B620} (2202) 237, hep-th/0106155.}

\lref\Vafawitten{ C.~Vafa and E.~Witten, {\it ``A One Loop Test of
String Duality''}, Nucl.\ Phys.\ {\bf B447} (1995) 261,
hep-th/9505053.}

\lref\SethiVW{ S.~Sethi, C.~Vafa and E.~Witten, {\it
``Constraints on Low-dimensional String Compactifications''},
Nucl.\ Phys.\ {\bf B480} (1996) 213, hep-th/9606122.}

\lref\HananyK{ A.~Hanany and B.~Kol, {\it ``On Orientifolds,
Discrete Torsion, Branes and M Theory''}, JHEP {\bf 0006} (2000)
013, hep-th/0003025.}

\lref\Ganor{ O.~J.~Ganor,{\it ``Compactification of Tensionless
String Theories''}, hep-th/9607092.}

\lref\DMtwo{ K.~Dasgupta and S.~Mukhi, {\it ``A Note on
Low-Dimensional String Compactifications''}, Phys.\ Lett.\ {\bf
B398} (1997) 285, hep-th/9612188.}

\lref\ShiuG{ B.~R.~Greene, K.~Schalm and G.~Shiu, {\it ``Warped
Compactifications in M and F Theory''}, Nucl.\ Phys.\ {\bf B584}
(2000) 480, hep-th/0004103.}

\lref\harmoni{ G.~W.~Gibbons and P.~J.~Ruback, {\it ``The Hidden
Symmetries Of Multicenter Metrics,''} Commun.\ Math.\ Phys.\ {\bf
115}, 267 (1988); N.~S.~Manton and B.~J.~Schroers, {\it ``Bundles
over Moduli Spaces and the Quantization Of BPS Monopoles,''}
Annals Phys.\  {\bf 225}, 290 (1993); A.~Sen, {\it ``Dyon -
Monopole Bound States, Selfdual Harmonic Forms on the Multi -
Monopole Moduli Space, and SL(2,Z) Invariance in String Theory,''}
Phys.\ Lett.\ B {\bf 329}, 217 (1994), hep-th/9402032.}

\lref\mesO{P.~Meessen and T.~Ortin, {\it ``An Sl(2,Z) Multiplet
of Nine-Dimensional Type II Supergravity Theories''}, Nucl.\
Phys.\ B {\bf 541} (1999) 195, hep-th/9806120; E. Bergshoeff, C.
Hull and T. Ortin, {\it ``Duality in the Type II Superstring
Effective Action''}, \np{451} {1995}{547}, hep-th/9504081;
S.~F.~Hassan, {\it ``T-duality, Space-Time Spinors and R-R fields
in Curved Backgrounds,''} Nucl.\ Phys.\ B {\bf 568}, 145 (2000),
hep-th/9907152.}

\lref\greenS{M.~B.~Green and J.~H.~Schwarz, {\it ``Superstring
Interactions,''} Nucl.\ Phys.\ B {\bf 218}, 43 (1983).}

\lref\SmitD{ B.~de Wit, D.~J.~Smit and N.~D.~Hari Dass, {\it
``Residual Supersymmetry Of Compactified D = 10 Supergravity''},
Nucl.\ Phys.\ {\bf B283} (1987) 165 (1987); N.~D.~Hari Dass, {\it
``A no-go theorem for de Sitter compactifications?,''} Mod.\
Phys.\ Lett.\ A {\bf 17}, 1001 (2002), hep-th/0205056.}

\lref\PapaDI{ S.~Ivanov and G.~Papadopoulos, {\it ``A No-Go
Theorem for String Warped Compactifications''}, Phys.\ Lett.{\bf
B497} (2001) 309, hep-th/0008232.}

\lref\DineSB{M.~Dine and N.~Seiberg, {\it ``Couplings and Scales
in Superstring Models''}, Phys.\ Rev.\ Lett.\  {\bf 55}, 366
(1985).}

\lref\olgi{O. DeWolfe and S. B. Giddings, {\it ``Scales and
Hierarchies in Warped Compactifications and Brane Worlds''},
hep-th/0208123.}

\lref\hellermanJ{S.~Hellerman, J.~McGreevy and B.~Williams, {\it
``Geometric Constructions of Non-Geometric String Theories''},
hep-th/0208174.}

\lref\WIP{K. Becker, M. Becker, K. Dasgupta, {\it Work in
Progress}.}

\lref\tatar{K.~Dasgupta, K.~h.~Oh, J.~Park and R.~Tatar, {\it
``Geometric Transition Versus Cascading Solution,''} JHEP {\bf
0201}, 031 (2002), hep-th/0110050.}

\lref\civ{F.~Cachazo, B.~Fiol, K.~A.~Intriligator, S.~Katz and
C.~Vafa, {\it ``A geometric unification of dualities,''} Nucl.\
Phys.\ B {\bf 628}, 3 (2002), hep-th/0110028.}

\lref\renata{K.~Dasgupta, C.~Herdeiro, S.~Hirano and R.~Kallosh,
{\it ``D3/D7 Inflationary Model and M-theory,''} Phys.\ Rev.\ D
{\bf 65}, 126002 (2002), hep-th/0203019; K.~Dasgupta, K.~h.~Oh,
J.~Park and R.~Tatar, {\it ``Geometric Transition Versus
Cascading Solution,''} JHEP {\bf 0201}, 031 (2002),
hep-th/0110050.}

\lref\carlos{R.~Kallosh, {\it ``N = 2 Supersymmetry and de Sitter
Space,''} hep-th/0109168; C.~Herdeiro, S.~Hirano and R.~Kallosh,
{\it ``String Theory and Hybrid Inflation / Acceleration,''} JHEP
{\bf 0112}, 027 (2001), hep-th/0110271.}

\lref\trivsham{S.~Kachru, R.~Kallosh, A.~ Linde, J.~Maldacena,
L.~ McAllister, S.~Trivedi, {\it Work in Progress}.}

\lref\polchinski{ J.~Polchinski, {\it ``String Theory. Vol. 2:
Superstring Theory And Beyond''}.}

\lref\taylor{E.~Cremmer, S.~Ferrara, L.~Girardello and A.~Van
Proeyen, {\it ``Yang-Mills Theories with Local Supersymmetry:
Lagrangian, Transformation Laws and Superhiggs Effect,''} Nucl.\
Phys.\ B {\bf 212}, 413 (1983); T.~R.~Taylor and C.~Vafa, {\it
``RR Flux on Calabi-Yau and Partial Supersymmetry Breaking,''}
Phys.\ Lett.\ B {\bf 474}, 130 (2000), hep-th/9912152.}

\lref\guko{S.~Gukov, C.~Vafa and E.~Witten, {\it ``CFT's from
Calabi-Yau Four-folds,''} Nucl.\ Phys.\ B {\bf 584}, 69 (2000)
[Erratum-ibid.\ B {\bf 608}, 477 (2001)], hep-th/9906070.}

\lref\hullwitten{ C.~M.~Hull and E.~Witten, {\it ``Supersymmetric
Sigma Models and the Heterotic String,''} Phys.\ Lett.\ B {\bf
160}, 398 (1985); A.~Sen, {\it ``Local Gauge And Lorentz
Invariance Of The Heterotic String Theory,''} Phys.\ Lett.\ B
{\bf 166}, 300 (1986); {\it ``The Heterotic String In Arbitrary
Background Field,''} Phys.\ Rev.\ D {\bf 32}, 2102 (1985); {\it
``Equations Of Motion For The Heterotic String Theory From The
Conformal Invariance Of The Sigma Model,''} Phys.\ Rev.\ Lett.\
{\bf 55}, 1846 (1985).}

\lref\dasmukhi{ K.~Dasgupta and S.~Mukhi, {\it ``F-theory at
Constant Coupling,''} Phys.\ Lett.\ B {\bf 385}, 125 (1996),
hep-th/9606044.}

\lref\vafasen{ C.~Vafa, {\it ``Evidence for F-Theory,''} Nucl.\
Phys.\ B {\bf 469}, 403 (1996), hep-th/9602022; A.~Sen, {\it
``F-theory and Orientifolds,''} Nucl.\ Phys.\ B {\bf 475}, 562
(1996), hep-th/9605150; T.~Banks, M.~R.~Douglas and N.~Seiberg,
{\it ``Probing F-theory with branes,''} Phys.\ Lett.\ B {\bf
387}, 278 (1996), hep-th/9605199.}

\lref\zwee{ M.~R.~Gaberdiel and B.~Zwiebach, {\it ``Exceptional
groups from open strings,''} Nucl.\ Phys.\ B {\bf 518}, 151
(1998), hep-th/9709013.}

\lref\callan{ C.~G.~Callan, E.~J.~Martinec, M.~J.~Perry and
D.~Friedan, {\it ``Strings In Background Fields,''} Nucl.\ Phys.\
B {\bf 262}, 593 (1985); A.~Sen, {\it ``Equations Of Motion For
The Heterotic String Theory From The Conformal Invariance Of The
Sigma Model,''} Phys.\ Rev.\ Lett.\  {\bf 55}, 1846 (1985); {\it
``The Heterotic String In Arbitrary Background Field,''} Phys.\
Rev.\ D {\bf 32}, 2102 (1985).}

\lref\WittenBS{ E.~Witten, {\it ``Toroidal Compactification
without Vector Structure''}, JHEP {\bf 9802} (1998) 006 (1998),
hep-th/9712028.}

\lref\carluest{ G.~L.~Cardoso, G.~Curio, G.~Dall'Agata, D.~Lust,
P.~Manousselis and G.~Zoupanos, {\it ``Non-K\"ahler String
Backgrounds and their Five Torsion Classes,''} hep-th/0211118.}

\lref\grifhar{ P.~ Griffiths and J.~ Harris {\it ``Principles of
Algebraic Geometry,''} Wiley Classics Library.}

\lref\GP{ E.~Goldstein and S.~Prokushkin, {\it ``Geometric Model
for Complex non-Kaehler Manifolds with SU(3) Structure,''}
hep-th/0212307.}

\lref\toappear{ K.~Becker, M.~Becker, K.~Dasgupta, P.~S.~Green,
..., {\it ``Work in Progress.''}}

\lref\MeessenQM{ P.~Meessen and T.~Ortin, {\it ``An Sl(2,Z)
Multiplet of Nine-Dimensional Type II Supergravity Theories''},
Nucl.\ Phys.\ B {\bf 541} (1999) 195, hep-th/9806120.}

\lref\HUT{ C.~M.~Hull and P.~K.~Townsend, {\it ``The Two Loop
Beta Function for Sigma Models with Torsion,''} Phys.\ Lett.\ B
{\bf 191}, 115 (1987); {\it ``World Sheet Supersymmetry and
Anomaly Cancellation in the Heterotic String,''} Phys.\ Lett.\ B
{\bf 178}, 187 (1986).}

\lref\SenJS{ A.~Sen, {\it ``Dynamics of Multiple Kaluza-Klein
Monopoles in M and String Theory''}, Adv.\ Theor.\ Math.\ Phys.\
{\bf 1} (1998) 115, hep-th/9707042. }

\lref\sav{ K.~Dasgupta, G.~Rajesh and S.~Sethi, {\it ``M theory,
orientifolds and G-flux,''} JHEP {\bf 9908}, 023 (1999),
hep-th/9908088.}

\lref\BeckerNN{ K.~Becker, M.~Becker, M.~Haack and J.~Louis, {\it
``Supersymmetry Breaking and $\alpha'$-Corrections to Flux
Induced  Potentials''}, JHEP {\bf 0206} (2002) 060,
hep-th/0204254.}

\lref\maeda{K.~i.~Maeda, {\it ``Attractor In A Superstring Model:
The Einstein Theory, The Friedmann Universe And Inflation,''}
Phys.\ Rev.\ D {\bf 35}, 471 (1987).}

\lref\rohmwit{R.~Rohm and E.~Witten, {\it ``The Antisymmetric
Tensor Field In Superstring Theory,''} Annals Phys.\  {\bf 170},
454 (1986).}

\lref\kg{S.~Gukov, S.~Kachru, X.~Liu, L.~McAllister, {\it To
Appear}.}

\lref\eva{E.~Silverstein,
{\it ``(A)dS backgrounds from asymmetric orientifolds,''} hep-th/0106209;
S.~Hellerman, J.~McGreevy and B.~Williams,
{\it ``Geometric constructions of nongeometric string theories,''}
hep-th/0208174;
A.~Dabholkar and C.~Hull,
{\it ``Duality twists, orbifolds, and fluxes,''} hep-th/0210209.}

\lref\renshamit{ S.~Kachru, R.~Kallosh, A.~Linde, S.~Trivedi,
{\it ``de-Sitter Vacua in String Theory,''} hep-th/0301240.}

\lref\bbdg{K.~Becker, M.~Becker, K.~Dasgupta and P.~S.~Green,
{\it ``Compactifications of heterotic theory on non-K\"ahler
complex
 manifolds.  I,''} hep-th/0301161.}

\lref\candle{P.~Candelas, G.~T.~Horowitz, A.~Strominger and E.~Witten,
{\it ``Vacuum Configurations For Superstrings,''}
Nucl.\ Phys.\ B {\bf 258}, 46 (1985);
A.~Strominger and E.~Witten,
{\it ``New Manifolds For Superstring Compactification,''}
Commun.\ Math.\ Phys.\  {\bf 101}, 341 (1985).}

\lref\weba{J.~Bagger and J.~Wess, {\it ``Supersymmetry and
Supergravity,''} Princeton University Press.}

\lref\becons{ M.~Becker and D.~Constantin,{\it ``A Note on Flux
Induced Superpotentials in String Theory ''}, hep-th/0210131.}

\lref\ccdl{G.~L.~Cardoso, G.~Curio, G.~Dall'Agata and D.~Luest,
{\it ``BPS Action And Superpotential For Heterotic String Compactifications
With Fluxes''}, hep-th/0306088.}

\lref\witsuper{E.~Witten, {\it ``New Issues in Manifolds of
$SU(3)$ Holonomy}, Nucl.\ Phys.\ B {\bf 268} (1986) 79.}

\lref\douglas{M.~R.~Douglas, {\it ``The Statistics of String/M
Theory Vacua''}, hep-th/0303194.}

\lref\toappear{ K.~Becker, M.~Becker, K.~Dasgupta, E.~Goldstein,
P.~S.~Green and S.~Prokushkin {\it ``Work in Progress.''}}

\lref\anna{A.~Fino and G.~Grantcharov,
{\it ``On some properties of the manifolds with skew-symmetric
torsion and holonomy SU(n) and Sp(n),''} math.dg/0302358.}

\lref\bg{K.~Behrndt and S.~Gukov, {\it ``Domain Walls and
Superpotentials from M theory on Calabi-Yau Three Folds''},
Nucl.\ Phys.\ B {\bf 580} (2000) 225, hep-th/0001082.}

\Title{\vbox{\hbox{hep-th/0304001} \hbox{UMD-PP-03-040,
SU-ITP-03/04}}} {\vbox{ \hbox{\centerline{Properties Of Heterotic
Vacua}} \vskip.05in \hbox{\centerline{From Superpotentials}}}}

\centerline{\bf Katrin Becker{$^1$}, Melanie Becker{$^2$}, Keshav
Dasgupta{$^3$}, Sergey Prokushkin{$^3$}} \vskip 0.2in \centerline
{ ${}^1$ Department of Physics, University of Utah, Salt Lake
City, UT 84112-0830}
\centerline{\tt katrin@physics.utah.edu} 
\centerline { ${}^2$ Department of Physics, University of
Maryland, College Park, MD 20742-4111}
\centerline{\tt melanieb@physics.umd.edu} 
\centerline {${}^3$ Department of Physics, Varian Lab, Stanford
University, Stanford CA 94305-4060} \centerline{\tt
keshav@itp.stanford.edu, prok@itp.stanford.edu}

\vskip.3in

\centerline{\bf Abstract}

\noindent We study the superpotential for the heterotic string
compactified on non-K\"ahler complex manifolds. We show that many
of the geometrical properties of these manifolds can be
understood from the proposed superpotential. In particular we
give an estimate of the radial modulus of these manifolds. We
also show, how the torsional constraints can be obtained from this
superpotential.

\Date{March, 2003}

\listtoc
\writetoc

\newsec{Introduction and Summary}

The heterotic string theory compactified on a Calabi-Yau
three-fold \candle\ preserving an ${\cal N} = 1$ supersymmetry in
four dimensions, has had great success in explaining many
interesting dynamics of nearly standard like models in string
theory (for a very recent discussion see e.g. \douglas). However,
in the absence of {\it rigid} vacua, there are many uncontrolled
moduli, that are not fixed, at least at the tree level. Very
generally, these moduli originate from the topological data of
the internal manifold. The K\"ahler structure moduli and the
complex structure moduli, classified by the hodge numbers
$h^{1,1}$ and $h^{2,1}$ of the internal manifold respectively,
give rise to a number of massless scalars in four dimensions in
addition to the scalars, that we get from other $p$-form fields.
As a result, the predictive power of string theory and ${\cal
M}$-theory is lost, as the expectation value of these moduli
fields determines the coupling constants of the standard model.
This pathology can be overcome, if we can lift these moduli, by
giving masses to these scalars. Since the radius of the internal
manifold is also a modulus, the size of this manifold cannot
be determined in a conventional Calabi-Yau compactification.
Considering now the fact that quantum effects favour the
limit of infinite radius, the radius of the manifold will eventually
become very big. This
is the so called Dine-Seiberg runaway problem \DineSB, which has
been one of the most important problems of string theory for a
long time. Luckily, in recent times we are making a remarkable
progress in this direction, by considering string theory and
${\cal M}$-theory compactifications with non-vanishing expectation
values for $p$-form fluxes. These compactifications are
generalizations of the conventional Calabi-Yau compactifications.

In the context of the recent developments in non-perturbative
string theory, compactifications with non-vanishing fluxes were
first found in \rBB, where ${\cal M}$-theory compactifications on
complex four-manifolds were described in detail. As was shown
therein, compactifications with fluxes, which preserve an ${\cal
N}=2$ supersymmetry in three dimensions can be obtained by
considering a more general type of compactification on a warped
background. The idea of having a warped background has appeared
in many earlier papers, even before the recent developments in
non-perturbative string theory were made, particularly in
\rstrom,\HULL,\hullwitten,\gates\ and \SmitD. However, in all
these compactifications precise models describing these
backgrounds were never computed very explicitly.

The warped backgrounds found in \rBB\ were later extended in two
different directions. The first interesting direction was pointed
out by \guko. These authors showed, that by switching on
four-form fluxes in ${\cal M}$-theory it is possible to generate a
superpotential, which freezes many of the moduli appearing in
these compactifications. In fact, all the complex structure
moduli and some K\"ahler structure moduli are fixed in the
process. This is an important progress, because as a result
string theory vacua with reduced moduli were generated even at
string tree level. The fact, that fluxes freeze the moduli fields
at tree level is particularly attractive, as it becomes easier to
perform concrete calculations.

The second direction, in which \rBB\ was generalized, was taken in
\sav. Choosing a particular four-manifold, which is also a
particular ${\cal F}$-theory vacuum at constant coupling, it was
shown in \sav, that after a series of U-duality transformations,
one can obtain a six-dimensional compactification of the $SO(32)$
heterotic string theory. This is particularly interesting from
the phenomenological point of view. This compactification is
again a warped compactification, as one would expect, but the six
manifold is now inherently non-K\"ahler. In the earlier
compactifications of \rBB, the warped manifolds were non-K\"ahler
but conformally Calabi-Yau. Similar conformally Calabi-Yau
compactifications can also be studied in the context of Type II
theories (see e.g. the third reference in \gates\ and references
therein). By choosing different four-folds and then performing a
series of U-duality transformations, many new heterotic and Type
II compactifications on non-K\"ahler manifolds can be generated
\kst,\beckerD,\kstt,\pktspto,\pktsptt\ and \bbdg. Such
compactifications are fascinating both, from the physics point of
view, as they have many properties in common with the standard
model and from the mathematical point of view, as many
mathematical aspects of these manifolds are still unexplored
territory. Some of these new mathematical aspects have been
pointed out recently in \carluest,\louisL,\GP\ and \bbdg. The
concrete manifolds described in many of these papers are
non-trivial $T^2$ fibrations over a four-dimensional Calabi-Yau
base \sav,\beckerD,\GP\ and \bbdg. Furthermore, these manifolds
have a vanishing Euler number and a vanishing first Chern class
\GP,\bbdg.

One immediate next step would be to combine these two directions,
i.e. one would like to compute the form of the superpotential for
compactifications of the heterotic string on non-K\"ahler complex
manifolds in order to understand, how the moduli fields get
frozen in this type of compactifications, as this is a rather
important question for particle phenomenology. The non-K\"ahler
manifolds, that we consider support three-form fluxes, that are
real (we call them ${\cal H}$). These fluxes will generate a
superpotential in the heterotic theory and as a consequence many
of the moduli fields will be frozen. The form of this
superpotential has been computed in \pktsptt\ and in \bbdg. It is the
goal of this paper to study the properties of this superpotential
and its effect on the moduli fields appearing in these
compactifications. As one would expect from the above
discussions, all the complex structure moduli and some of the
K\"ahler structure moduli are fixed in the process. An important
K\"ahler structure moduli, that is frozen at tree level is the
radial modulus. This has been shown already in \bbdg\ and we
shall see this here in much more detail. An immediate consequence
of this is now apparent: there is no Dine-Seiberg runaway
behavior for the radius and therefore the notion of
compactification makes perfect sense, as the internal manifold
will have a definitive size. However, this is not enough. We have
to see whether supergravity analysis is also valid in
four-dimensions, so that explicit calculations can be done. This
would imply, that the internal six-manifold should have a large
overall volume. We will show that 
even though the $T^2$ fiber has a volume of order
$\alpha'$, the base can be made large enough, so that the total
volume is large. But there is a subtletly related to the topology
change which in fact hinders any nice supergravity description for the 
kind of background that we study. We will discuss this important issue later. 

This paper is organized as follows: In section 2 we give a brief
review of the earlier works on non-K\"ahler spaces. We show, how
these manifolds can be realized directly in the heterotic theory
without using T-duality arguments to the Type IIB theory \bbdg. In
section 3 we discuss the stabilization of the radial modulus.
This aspect has been partially discussed in \bbdg. Here we will
give a fuller picture of the potential, that fixes this modulus
and give a numerical estimate for the radius. To obtain this
estimate, we have made some simplifying assumptions. In section 4
we show, that the estimate done in section 3 is not too far from
what we obtain in a more realistic scenario. In section 5 we
calculate an additional contribution to the heterotic
superpotential and show, how the torsional constraints can be
derived from the complete superpotential. All the contributions to
the superpotential, that we have computed to this point are
perturbative. Section 6 is dedicated to discussions and
conclusions. In particular, we discuss the origin of
non-perturbative contributions to the superpotential and
applications of this type of compactifications in other possible
scenarios, such as cosmology.

\newsec{Brief Review of Torsional Backgrounds}

Here is a lightning review of earlier works on non-K\"ahler spaces
\sav,\beckerD,\GP,\bbdg. The readers are however advised to go
through these references, as we shall constantly be referring to
them. The non-K\"ahler manifolds, that we study here are all
six-dimensional spaces of the form \eqn\nkmet{ds^2 = \Delta_1^2~
ds^2_{CY} + \Delta_2^2 ~ |dz^3 + \alpha dz^1 + \beta dz^2|^2,} where
$\Delta_i=\Delta_i (|z_1|,|z_2|)$ are the warp factors and $\alpha$,
$\beta$ depend on $z^i$ and ${\bar z}^j$, the coordinates on the
internal space. The four-dimensional Calabi-Yau base is described
by $z^1$ and $z^2$. For the examples studied earlier in
\sav,\beckerD,\bbdg\ , these functions were \eqn\earlywor{\alpha
= 2i~ {\bar z}^2, ~~~\beta = -(4 + 2i) ~{\bar z}^1, ~~~
\Delta_1^2 \equiv \Delta^2 = c_{\rm o} + \psi(|z^1|,|z^2|), ~~~\Delta_2 = 1,}
where $c_{\rm o}$ is a
constant, and $\psi \to 0$ when size of the manifold becomes
infinite. There is also a background three-form , which is {\it
real} and anomaly free and serves as the torsion for the
underlying space. The dilaton is {\it not} constant and is
related to the warp factor. The background supports a modified
connection $\tilde\omega$ instead of the usual torsion-free
connection $\omega_o$. In fact, the torsion ${\cal T}$ is
proportional to \eqn\torsi{{\cal T} = \omega_o - \tilde\omega,}
which in particular is also the measure of the real three-form in
this background because we demand ${\cal T}$ to be covariant and
the ``contorsion'' tensor to be identified with it \HULL,\bbdg.
This identification of contorsion tensor to the heterotic
three-form actually has roots in the sigma model description of
the heterotic string propagating on these manifolds. Consider the
heterotic string with a gauge bundle $A_\mu$. If we define the
contorsion tensor $\kappa$ to satisfy \eqn\contor{ \kappa_\mu =
\omega_{o\mu}^{~~ab}\sigma^{ab} - A_\mu^{AB} T^{AB} + {\cal
O}(\alpha'),} where $\sigma,T$ are the generators of holonomy and
gauge groups respectively, then the theory becomes anomaly free
with an almost vanishing two-loop sigma model beta function.

There is however a subtlety here. The above identification,
though appears so natural, creates a problem, which is the
following. In the relation \contor, if we have an exact equality,
with vanishing terms of order ${\cal O}(\alpha')$, then the
two-loop beta function would cancel exactly. In that case we will
have {\it no} warped solution and the manifold will tend to go
back to the usual Calabi-Yau compactification. On the other hand,
having an ${\cal O}(\alpha')$ term would mean, that we have a
non-zero beta function and therefore, these compactifications are
{\it not} solutions of the string equations of motions. Either
way is disastrous, unless we find a way out, that could save the
day.

The resolution to this problem comes from the fact, that these
manifolds are in fact {\it rigid} and therefore, they do not have
an arbitrary size. Thus, even though we allow \contor\ with
non-vanishing terms at order $\alpha'$, the two-loop beta
function can become zero {\it only}, when the manifold attains a
definite size. For any arbitrary size the beta function is
non-zero and therefore our manifolds are not a solution of the
equations of motion (a similar argument goes through for all the
other moduli). Happily, as shown in \bbdg\ all the complex
structure moduli, some K\"ahler structure moduli, in particular,
the radial modulus do get stabilized in these compactifications
at tree level. The remaining K\"ahler moduli would also get
stabilized, if we incorporate quantum effects (see \BeckerNN).

\bigskip

\centerline{\epsfbox{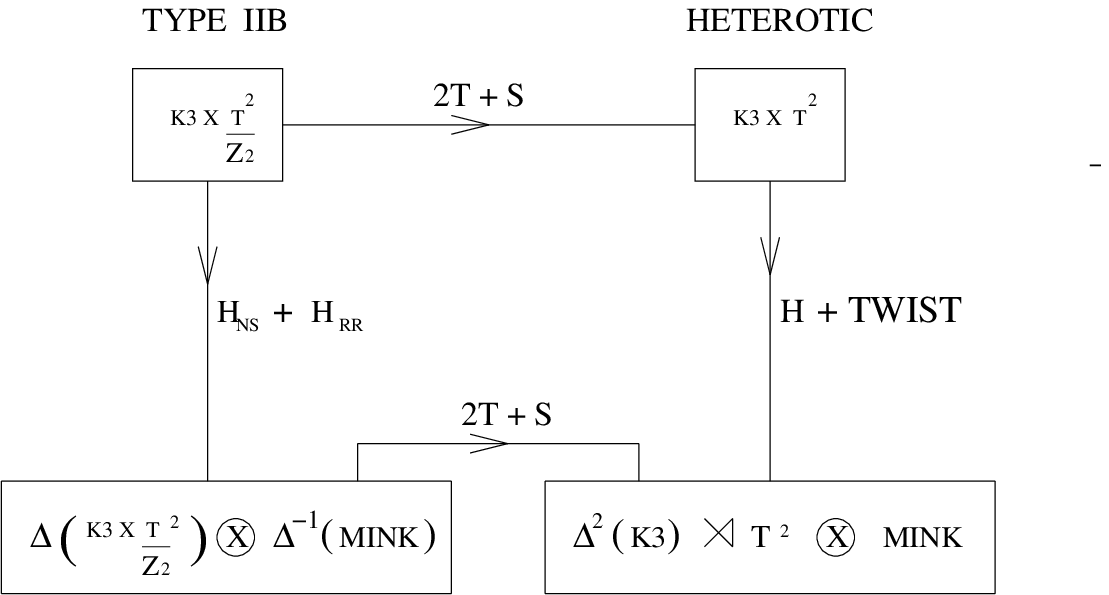}}\nobreak \centerline{{\bf Fig. 1}:
{\it The mapping of the Type IIB model to the heterotic model}}
\centerline{\it via two T-dualities and one S-duality.}

\bigskip

Before we go into discussing more details on these
compactifications, we should point out the fact, that given a
heterotic background, switching on a three-form will not, in
general, convert this to a non-K\"ahler manifold. In fact, this
is clear from the figure above.

We start with a warped background in the Type IIB theory on $K3
\times T^2/Z_2$ with fluxes, and through a set of U-duality
transformations we get the non-K\"ahler manifold discussed in
\sav,\beckerD,\bbdg. Observe, that in this process we actually
have a topology change, because on the heterotic side we go from
a torus $T^2$ to a fibered torus, having no one-cycle. Therefore,
we need more than a three-form background to fully realize the
non-K\"ahler spaces. We will dwell on this issue in the next
section.

There is also an ${\cal F}$-theory picture, from which the
construction of the heterotic manifold is rather straightforward.
This is ${\cal F}$-theory at a constant coupling, where the
elliptic curves degenerate to \eqn\ellip{y^2 = x^3 + a \phi^2 x +
\phi^3,} where $\phi = \phi(z)$ is an arbitrary polynomial of
degree 4 and $z$ is the coordinate of the $P^1$ base. In fact,
this is ${\cal F}$ theory on $K3 \times K3$, where one of the
$K3$ has degenerated to the $Z_2$ orbifold point. Under suitable
rescaling of the above curve \ellip, one can easily show, that at
a given orbifold point we have \vafasen\ \eqn\wehave{Y^2 = X^3 +
\a X z^2 + z^3,} where $X,Y$ can be derived by knowing $x,y,z$.
{}From Tate's algorithm we can see the appearance of a $D_4$
singularity at that point.

\newsec{Superpotential and Radial Modulus}

In a recent paper \bbdg\ we showed, how all the complex structure
moduli, some K\"ahler structure moduli and in particular the
radial modulus are determined at tree level by switching on
three-form fluxes in compactifications of the heterotic string on
non-K\"ahler complex six-dimensional manifolds. The basic idea
is, that a superpotential is induced by the fluxes. This gives
masses to most of the moduli. The superpotential takes the form
\eqn\superpot{
 W_{het} = \int G \wedge \Omega,}
where $G$ is a three-form and $\Omega$ is the holomorphic
(3,0)-form of the internal six-dimensional manifold. In the
following we would like to determine what $G$ is. In the usual
case where there is no torsion, $G$ is the {\it real} three-form
of the heterotic theory \becons. We still have the real
three-form in the presence of torsion (the torsion is actually
identified with this real form), but as discussed in \bbdg, there
is another choice for the three-form $G$ appearing in the above
superpotential, that is needed for non-K\"ahler internal
manifolds. This three-form $G$ is again anomaly free and gauge
invariant and satisfies the equation
\eqn\threeform{ G = d{\cal B} + \alpha'\left[\Omega_3\left(\omega_o
-{1\o 2} {\tilde G}\right) - \Omega_3({\cal A})\right],}
where $\Omega_3({\cal A})={\rm Tr}~({\cal A} \wedge F - {1 \over
3} {\cal A} \wedge {\cal A} \wedge {\cal A})$ is the Chern-Simons
term for the gauge field ${\cal A}$ and $\Omega_3(\omega_o)$ is
the Chern-Simons term for the torsion free spin-connection
$\omega_o$ (the trace will now be in the fundamental
representation whereas the trace above was for the adjoint
representation), while ${\cal B}$ is the usual two-form potential
of the heterotic theory. We have also defined ${\tilde G}$ as the
one-form created out of three-form $G$ using vielbeins $e^a_i$ as
${\tilde G}_i^{~ab}=G_{ijk}e^{aj}e^{bk}$. We see, that $G$
appears on both sides of the above expression and therefore we
need to solve iteratively this equation in order to determine
$G$. For the case considered here we can do this order by order
in $\alpha'$. The equation to be solved is
\eqn\cubic{ G + {\alpha'\o 2}{\rm tr} \left(\omega_o \wedge {\cal
R}_{\tilde G} + {\tilde G} \wedge {\cal R}_{\omega_o} - {1\o 2}
{\tilde G}\wedge {\cal R}_{\tilde G}\right) = d{\cal B} +
\alpha'\left(\Omega_3(\omega_o) - \Omega_3({\cal A})\right),}
where we have introduced the curvature polynomials ${\cal
R}_{\tilde G}$ and ${\cal R}_{\omega_o}$ as \eqn\defofcurv{ {\cal
R}_{\tilde G}=d \tilde G - {1\o 3}\tilde G \wedge \tilde G,
~~~{\rm and} ~~~{\cal R}_{\omega_o} = d\omega_o + {2\o 3}\omega_o
\wedge \omega_o.} To the lowest order in $\alpha'$ we can ignore
the contributions from $d{\tilde G}$ \cubic, because they are of
$O(\alpha'^2)$. We will also ignore the contributions from
$d\omega_o$, because they are higher derivatives in the
vielbeins. The above formula reduces to the usual heterotic
three-form equation in the absence of torsion and the
superpotential becomes the superpotential computed in \becons, as
can be easily seen. Now if we denote the size of the internal
manifold as $t$ (we shall take $t$ to be a function of all the
spatial coordinates), we obtain from \cubic\ a cubic equation,
which takes the generic form
\eqn\cubicnow{ h^3 + ph + q =0, \qquad {\rm with} \qquad G_{ijk} =
h~ {\cal C}_{ijk}, \qquad {\rm and} \qquad g_{ij} = t~g^{\rm
o}_{ij},}
for {\it every} component of the three-form $G$. Here $\cal C$ is
a constant antisymmetric tensor in six-dimensions, whose
contractions are done with respect to the metric $g^{\rm o}_{ij}$.
And $g^o_{ij}$ is chosen to be constant locally, so that we can
ignore the twist of the fiber\foot{In other words, $\alpha$ and
$\beta$ in \nkmet\ are constants locally.}. Also, we can relax
the condition on ${\cal C}$ a little bit. What we actually require
is, that ${\cal C}$ should be at least anti-symmetric in two of
its indices. However, for all the calculations below we will only
use the complete antisymmetric part\foot{An alternative way to
think about this is to regard ${\tilde G}_i^{~ab}$ in the same
way as $\omega_{oi}^{~~ab}$. Thus, we define ${\tilde G}_i^{~ab}
= t^{-1} h~ {\cal C}_i^{~ab}$ and, therefore $G_{[ijk]} = {\tilde
G}_{[i}^{~ab}~e_{|a| j}~e_{|b| k]} = h~{\cal C}_{[ijk]}$. Only
the anti-symmetric part of ${\cal C}$ will be relevant.} of
${\cal C}$ in analogy to the torsion-free spin-connection
$\omega_o$. Here we will consider again only the antisymmetric
part, unless mentioned otherwise. Another point to note is the
choice of metric in \cubicnow. Our assumptions for ${\cal C}$ and
the metric can therefore be summarized as \eqn\assumpta{{\cal
C}_{[ijk]} = \epsilon_{ijk}, ~~ \qquad ~~ \Delta_1=\Delta_2 = 1,} where
$\Delta_i$ are the warp factors in \nkmet. This will simplify the
calculations done below. In the next section we will consider the
case, where the warp-factor is introduced back as $\Delta_1 = \Delta_2 =
\Delta$. We will, however, not go in much details for the case $\Delta_1 =
\Delta, \Delta_2 = 1$ which is a little subtle and needs a more detailed
analysis than what will be presented here.
The calculation in full generality is not
too different from the simple example, that we are considering
herein. We will be using the definitions of $p,q$ and $f$
as\foot{We will absorb the traces of the holonomy matrices in the
definition of $t$ for simplicity (as in \bbdg). However, we will
soon consider the case, where we keep all these dependences
explicitly.} \eqn\defoff{ p= {t^3 \o \a'},~~~~~q = -{f t^3\o
\a'},~~~~{\rm and}~~~ f = \left(d{\cal B} +
\alpha'\Omega_3(\omega_o) - \alpha'\Omega_3(A)\right)_{ijk}
\epsilon^{ijk}.} The first equation in \cubicnow\ has three
roots. One of them is real and the other two are complex
conjugates of each other. The real root appears in the
supersymmetry transformation of the low energy effective action
of the heterotic string and satisfies the torsional equations of
\HULL, \rstrom\ and  \beckerD. But for the construction of the
superpotential the real root is not enough, as we will explain in
the next paragraph.

The real solution fails to cover many interesting aspects of the
non-K\"ahler geometry. So for example, one particular important
aspect of non-K\"ahler manifolds, that was studied in \bbdg\ is
topology change. We start with a complex three manifold of the
form $K3 \times T^2$ in the heterotic theory and then switch on a
three-form flux. The final picture is, that we get a non-K\"ahler
complex three-fold, whose first Betti number, $b_1$, is zero.
Therefore, a transition from $b_1 = 2 \to b_1 = 0$ (see fig 1)
has been performed. In the usual perturbative analysis it is
difficult to see, how such a transition could take place by
switching on a torsion three-form\foot{We could also have $b_1 =
2 \to b_1 = 1$, but that would be for a slightly different choice
of non-K\"ahler manifold. Details on this have been discussed in \GP,
\bbdg.}. One needs an additional non-trivial {\it twist} in the
geometry to achieve such a topology change. Therefore, we need
both: a three-form background (i.e torsion) and a twist. The twist
is proportional to the antisymmetrized
spin-connection, because that is the only
gravitational degree of freedom generating such a change. We can
combine these two to form a {\it complex} three-form. This is
precisely what we get by solving the cubic equation above!

However, an immediate question would be: why a complex three-form
instead of a real one? Of course, we cannot have any arbitrary
combination of three-forms, because this would be inconsistent
with the dynamics of the heterotic theory, let alone the fact,
that it will be anomalous. There is however, a deeper reason of
why we have a complex three-form. This is related to the fact,
that the complex three-form is  compatible with the T-dual Type
IIB framework. In Type IIB theory we can have a complex
superpotential given in terms of NS-NS and R-R three-forms
$H_{NS}$ and $H_{RR}$ respectively as \eqn\iibsuper{W = \int
{\Big (}H_{RR} + \phi H_{NS} + {i\over g_s} H_{NS}{\Big )}\wedge
\Omega,} where $\phi$ is the axion and $g_s$ is the Type IIB
coupling constant related to the dilaton. For the simplest case,
where we take a vanishing axion-dilaton, the Type IIB
superpotential is given simply in terms of a complex three-form
$G_3 = H_{RR} + i H_{NS}$. We can go to the heterotic theory by
making two T-dualities and an S-duality, as shown in \sav\ and
\beckerD. Then $H_{RR}$ becomes the real heterotic
three-form\foot{To see the Chern-Simons part of the real
three-form ${\cal H}$, one has to carefully study the
singularities in the dual ${\cal M}$-theory setup. The localized
fluxes at the singularities and the non-zero curvature at those
points conspire precisely to give the Chern-Simons part. These
calculations have been done in detail in \bbdg\ and therefore we
refer the reader to this paper for more details.} and $H_{NS}$
becomes the spin-connection \kstt. These two fields combine in
the heterotic theory to give a complex superpotential. This is
again, what we get from our cubic equation.

The skeptic reader might still ask, whether such a complex
superpotential could be obtained directly from the supersymmetry
transformation rules or equivalently, from the lagrangian of the
heterotic theory. Since the whole heterotic dynamics can be
described in the T-dual Type IIB framework, where there exist a
complex three-form $G_3$, we could as well write our heterotic
lagrangian in terms of $G$, by combining the spin-connection part
and the real three-form part. This is an obvious straightforward
exercise, that can be easily performed.

We can be a bit more precise here. Let us consider the
usual heterotic lagrangian. In terms of the conventions that
we followed here, this is given as (we use the notations of \polchinski)
\eqn\hetlag{S= {1\o \kappa_{10}^2} \int d^{10}x~\sqrt{g} ~e^{-2\phi}
\left[ R + 4 |\del \phi|^2 - {1\o 2} |f|^2 + {\kappa_{10}^2\o g_{10}^2}
{\rm Tr} |F|^2 + {\cal O}(\alpha'^2) \right]}
where $\kappa_{10}$ and $g_{10}$ are defined in \polchinski, $F = d{\cal A}
+ {\rm Tr} {\cal A} \wedge {\cal A}$ and $f$ is given in \defoff. We can
rewrite the above lagrangian alternatively, to all orders in $\alpha'$,  as
\eqn\hetalter{S = \int_{{\cal M}_6}~e^{-2\phi} {\Big [} 2 |G|^2 +
{\rm Tr} |F|^2 + \sum_{m,n,p} a_{mnp} G^m F^n R^p {\Big ]} - 
{1\o \kappa_{4}^2} \int d^{4}x~\sqrt{g_4} ~e^{-2\phi}
|\del \phi|^2 + ....}
where the interaction terms are to be contracted properly to form scalars. The
coefficient $a_{mnp}$ is a constant (upto powers of dilaton) and we have 
denoted the non-K\"ahler manifold as ${\cal M}_6$. 
Observe that in the above
lagrangian we haven't yet defined $G$. We require $G$ to satisfy the
following conditions:

\noindent (a) It should be complex.

\noindent (b) It should be anomaly free and gauge invariant, and

\noindent (c) It should be locally represented as
\eqn\glocal{G = a~({\cal H} + ...) +  ib~(\omega_o + ...),} where
${\cal H}$ is the real three-form of the heterotic theory (the real root
of the anomaly equation), $a~{\rm and}~b$ are
arbitrary constants and the dotted terms will be estimated soon.

In the following analysis we will show that the complex root of the cubic
equation does satisfy all the above
three conditions, modulo possible geometric terms (and hence gauge invariant)
that could in principle
contribute to the imaginary part of the three-form $G$.\foot{We thank
Xiao Liu for discussion on this aspect.} We will ignore these contributions
for the time being and only mention them later.

Let us now study the solutions of
\cubicnow\ carefully. The three roots of the cubic equation
\cubicnow\ can be written in terms of $p$ and $q$. We define two
variables $A$ and $B$, that are functions of $p,q$, such that the
roots of the cubic equations are \eqn\roots{A + B, ~~~~ -{1\over
2}(A + B) \pm i {\sqrt{3}\over 2} ( A - B).} The variables $A,B$
are defined in \bbdg\ and are real. Therefore, the real root of
the cubic equation is $A + B$. This is in fact the heterotic three-form
that appears in the lagrangian. The series expansion
of this three-form in terms of powers of $p,q$ can be written as
\eqn\Aseries{A + B = -{q\over p} + {q^3\over p^4} - {3q^5\over
p^7} + {12 q^7\over p^{10}} - {55 q^9\over p^{13}} + {\cal
O}(q^{11}).} For the analysis done in \bbdg, we have taken the
definitions of $p,q$ appearing in
\defoff. But these are only valid, when we ignore the contributions coming
from the spin-connection $\omega_o$. When we take these into
account, the equations do get a little more complicated, as we
shall discuss in the next section. For the time being we shall
discuss a simple toy example, where we take the spin-connection
as $\omega_{o[mnp]} = \omega_o~\epsilon_{mnp}$. Without loss of
generality, and keeping terms only to the linear order in
$\omega_o$, we change the definitions of $p$ and $q$ to
\eqn\newpandq{p = {t^3\o \alpha'} - c~f~ \omega_o, ~~~~~ q = -{f
t^3\o \alpha'} + b~f^2 \omega_o,} where $c$ and $b$ are constants.
The above form of $p,q$ can be derived for the realistic case,
where we consider all the components of $G_{ijk}$ and we shall do
this in section 4. Furthermore, we are also ignoring possible
constant shifts in $p,q$ for simplicity. As discussed in \bbdg,
the expansion in powers of ${1\over p}$ is still a reasonable
thing to do, because this is a small quantity, as long as the size
of the three manifold, $t$, is a large number. The above
expansion can actually be terminated at order $\alpha'$, because
we have ignored contributions from $dG$ and $d\omega_o$, as these
contributions are higher orders in $\alpha'$ and higher derivatives in
$t$ respectively.
Doing this and
calling the real solution as ${\cal H}$, we obtain
\eqn\realh{{\cal H} = f - {\alpha'f^3\over t^3} +
{\alpha'\omega_o (c-b)f^2 \over t^3} + {\cal O}(\alpha'^2).} We
make two observations here. First, all the terms in the above
expansion are dimensionally the same as $f$. For the  case
that ${\cal H}$ is given by $f$
\defoff,
as in the usual Calabi-Yau compactifications, terms of ${\cal
O}({1\over t}) \to 0$, which means, that the radius of the
manifold goes off to infinity. This is precisely the Dine-Seiberg
runaway problem \DineSB, which does not appear in this type of
compactifications. For compactifications on non-K\"ahler complex
manifolds the real three-form gets modified to the value given in
\realh\ and the size of the radius is then finite. Secondly, note
that for the usual case when we have no $G$ dependences in the
Chern-Simons part of the three-form in \threeform, this would
have remained unaffected by scalings of the metric (because the
torsion-free spin connection $\omega_{o\mu}^{ab}$ is unaffected by
scalings of the vielbeins) and therefore the radius would {\it
not} have been stabilised. In the presence of $G$ in the
Chern-Simons part of \threeform\ the two sides of the equation
\threeform\ scale differently and therefore the radial modulus is
fixed. This is one of the basic advantages of torsional
backgrounds.

Let us now discuss the complex solutions. We see from the choice
of the roots \roots, that we need the expansion for $A - B$. This
is given in terms of the ${1\over p}$ expansion as
\eqn\Bseries{(A - B) = {2\over \sqrt{3}}{\Bigg [}\sqrt{p} +
{3\over 8} {q^2\over p^{5/2}} - {105\over 128}{q^4\over p^{11/2}}
+ {3003\over 1024}{q^6\over p^{17/2}} -{415701\over
32768}{q^8\over p^{23/2}}{\Bigg ]} + {\cal O}(q^{10}).} As we see
in \roots, this is just a part of the complex roots, as there is
a contribution from the real root ${\cal H}$. Again, we will keep
the expansion to order $\alpha'$. We call the complex roots as
$G$, with $G$ now given as \eqn\comroot{G = -{1\over 2}{\Bigg (}f
- {\alpha'f^3\over t^3} + {\alpha'\omega_o (c-b)f^2 \over
t^3}{\Bigg )} \pm i {\Bigg (}\sqrt{t^3\over \alpha'} + {3
f^2\over 8} \sqrt{\alpha'\over t^3} - {\omega_o~c~ f\over
2}\sqrt{\alpha'\over t^3}{\Bigg )} + {\cal O}(\alpha'^{3/2}),}
where to this order we do not see the effect of the constant $b$
in the imaginary part.\foot{This constant starts affecting the
expansion at the next order ${\sqrt{\alpha'^{3}\o t^{9}}}$ as
${3\o 8}(2.5 c - 2 b) f^3 -{105\o 128}f^4$.} An immediate
disconcerting thing about the above expansion might be the fact,
that the spin connection $\omega_o$ appears with a coefficient
$f$ in the imaginary part of $G$, as this is not expected from
T-duality arguments. However, this is an illusion. As has been
shown in \bbdg\ and as we shall see in a moment in more detail,
the size of the internal manifold is fixed by the choice of
background $f$ (see equation $3.31$ below). Using the relation
between $f$ and $t$ in \comroot, we can show, that the complex
root locally takes the form \eqn\comlook{ G = -{{\cal H}\over 2}
\mp i (\beta~ \omega_o + ...),} where $\beta$ is a pure constant
and the dotted terms involve contributions from the radial
modulus $t$, that in general could be functions of $f$ as well as $\omega_o$.
The above equation \comlook\ is, what we expected from the T-dual
Type IIB framework, because under T and S dualities the
three-form tensor fields $H_{RR}$ and $H_{NS}$ of the Type IIB
theory transform into the {\it real} heterotic three-form ${\cal
H}$ and the spin-connection respectively\foot{An important point to note
here is the following: Incorporating higher order polynomials into the
cubic equation (by putting in the values of $dG$ iteratively), the complex
root will change. In that case the real part of $G$ will shift from
the value ${\cal H}$ by additive factors as discussed in \glocal. To the
lowest order in $\alpha'$ (which gives us the cubic equation) we do not see
the shift.}.
However, T-duality
rules are only derived to the lowest order in $\alpha'$, which is
why we have performed our calculations directly in the heterotic
theory, instead of in the T-dual Type IIB theory. We shall
nevertheless use T-duality arguments from time to time for
comparison. Notice, that we can write the imaginary part of
\comroot\ as an {\it effective} spin-connection $\omega_{eff}$.
When we fix the radius of our manifold to a specific value, the
effective spin connection locally takes the form
\eqn\omegeff{\omega_{eff} = \omega_o + \gamma |f| + {\cal
O}(\omega_o^2, |f|^2),} where $\gamma$ is a constant related to
$\beta$ (we will soon give an estimate of this) and $|f|$ is the
background expectation value of $f$ given in \defoff. This
effective connection is {\it not} related to the modified
connection\foot{The modified connection, as derived earlier in
\bbdg, is $\tilde\omega = \omega_o - {1\o 2} {\cal H}$.}
 ${\tilde\omega}$ for the non-K\"ahler manifolds
and therefore shouldn't be confused with it. In fact this could
get contribution from other covariant terms briefly alluded to earlier.
Also we shall henceforth write
$\omega_{eff}$ as $\omega$, unless mentioned otherwise.

{}The complex three-form $G$, that we have computed in \comlook,
is what one would have expected from naive T-duality arguments,
except that there is an overall sign difference, an extra factor
of one half and the constant $\beta$ in front of the spin
connection, that originates from higher order $\alpha'$ effects
and thus cannot be seen by T-duality arguments. The overall factor
is not very important, because it can be easily absorbed into the
definition of the holomorphic (3,0) form $\Omega$. But the sign
is important. In fact, the sign in the above equation is not
difficult to explain. In the usual Type IIB picture there exists
a {\it perturbative} action of S-duality, that changes both
three-forms $H_{NS}$ and $H_{RR}$ by a sign, without changing any
other fields. This is the $SL(2, \IZ)$ operation
\eqn\sltwo{\pmatrix{-1&0 \cr 0&-1}.} After performing this
operation, we
 are left with the Type IIB three-form $G_3 = - H_{RR} - i H_{NS}$,
which under naive T and S dualities will give us precisely
\comlook. The way we have derived this form of $G$, is only valid
locally, because of the choice of the background \cubicnow\ and
\assumpta. But this local form of $G$ is consistent with, what
one would expect from T- and S-dualities, as we saw above. Taking
these considerations into account, our result for the
superpotential is of the form \eqn\suppot{W_{het} = \int ({\cal H}
+ i \beta~ \omega) \wedge \Omega.} For compactifications on
manifolds without torsion, where the Dine-Seiberg runaway problem
appears \DineSB, the complex three-form field becomes $G = f \pm i
\infty$ and therefore the imaginary part decouples from the path
integral. In this case all the cubic roots give the same result.
However, in the presence of torsion there is a splitting and
three different solutions appear. Notice also, that the complex
roots do not satisfy the torsional constraints as expected
(torsional constraints being real). This will be discussed in
section 5 in detail.

\bigskip

\centerline{\epsfbox{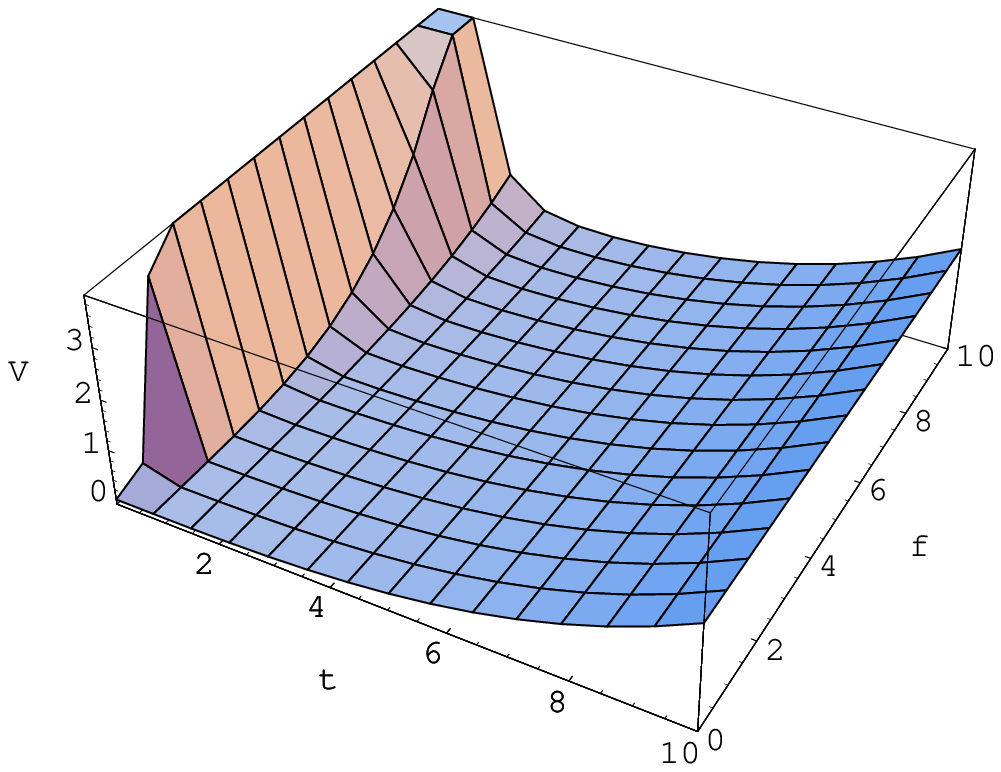}}\nobreak {\bf Fig. 2}: {\it
This is the form of the radial potential, taking the imaginary
part of the three-}{\it form into account. Along the x-axis we
have represented the radius $t$, along the y-axis the value of $f$
and along the z-axis}~{\it the potential $V(t,f)$. We have also
scaled down $V(t,f)$ }{\it by a factor of 500. Observe, that for a 
fixed value of $f$, the potential has a minimum.}
\bigskip

Let us now discuss the radial modulus stabilization. This issue
has already been addressed in some detail in \bbdg. However, in
\bbdg\ the potential was computed solely from the imaginary part
of the three-form. The result was shown to be approximately (see
fig. 2) \eqn\radmod{t = 0.722~ (\alpha' |f|^2)^{1/3},} where
$|f|$ is the expectation value of the background $f$ field. The
above value would shift a bit, if we include higher order
$\alpha'$ corrections to the three-form. If we now incorporate
the contribution coming from the real part of the three-form, the
potential for the radial modulus becomes \eqn\potrad{V(t) =
{t^3\over \alpha'} - {2\alpha' f^4\over t^3} + {7 \alpha'^2 f^6
\over  t^6},} where we keep terms to order $\alpha'^2$. Observe,
that the spin-connection dependent term cannot contribute to the
potential. The above potential fixes the radius of our manifold
to be (see fig. 3) \eqn\radnow{t = 1.288~(\alpha' |f|^2)^{1/3},}
which is a little larger, than the radius calculated in \radmod.
As we shall mention below, the actual value for the radial
modulus is smaller, than the result given in \radnow, as there
are other effects, that we need to take into account. As an
aside, it is interesting to note, that the real part of the
three-form fixes the value of the radius to be proportional to
$(\alpha' |f|^2)^{1/3}$, but gives an imaginary answer, when we
go to the next order.

\bigskip

\centerline {\epsfbox{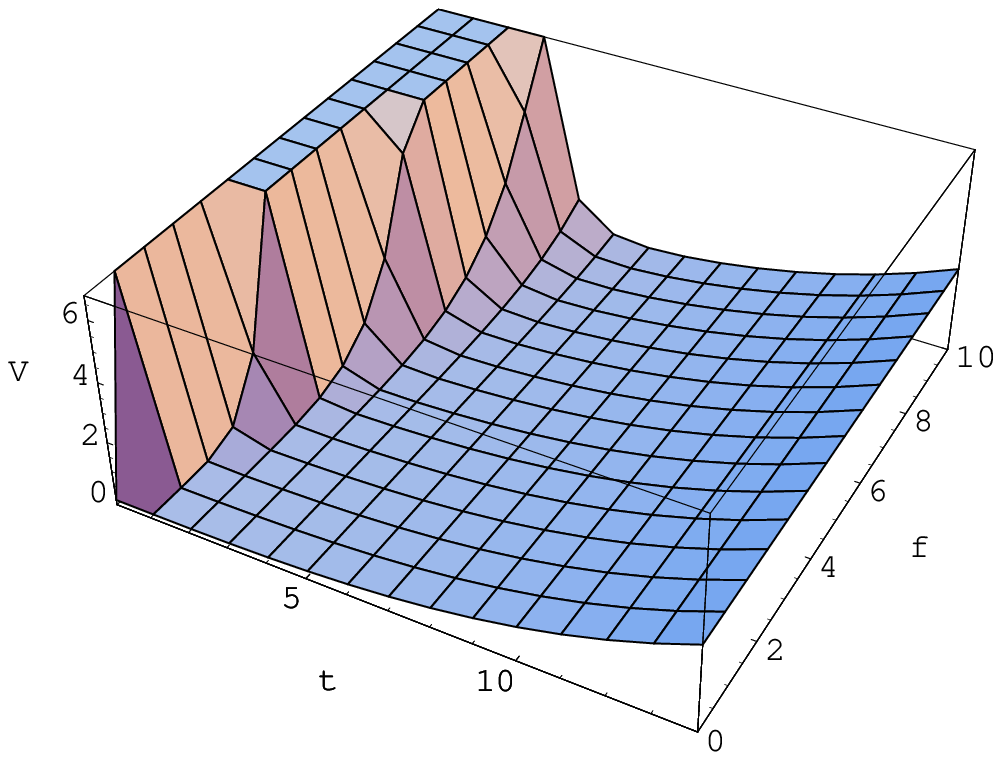}}\nobreak {{\bf Fig. 3}: {\it
The radial potential considering both the imaginary and the real
part of~}}{{\it the three-form. There is again a minimum, but now
shifted to a slightly larger~}}{\it value. Here we have scaled
down the potential $V(t,f)$ by a factor of 1000.}
\bigskip

In deriving \radnow\ we have not taken into account the fact, that
the radius $t$ also depends on the representation of the holonomy
group. This fact was partially alluded to in a previous paper
\bbdg. If we call the original radius as $t_o$, then $t$ is
related to $t_o$ by the following relation
\eqn\trel{t^{3}=t^{3}_o~{\rm tr}(M^{ab}M^{cd}M^{ef})~ \alpha^{ab}
\alpha^{cd} \alpha^{ef},} where $M^{ab}$ are the representation
of the holonomy group and $\alpha^{ab}$ depend on the background
three-form field and vielbein, as we describe in the following
(this has been discussed earlier in \bbdg). The one-form ${\tilde
G}_\mu^{~ab}$, which is relevant in the context of the heterotic
theory, when constructing the Chern-Simons form (see \threeform),
is given in terms of vielbeins $e^a_\mu$ as \eqn\onefor{{\tilde
G}_\mu^{~ab}dx^\mu = 2 {G}_{\mu\nu\rho} e^{\nu[a}e^{b]\rho}
dx^\mu \equiv t^{-1} h \alpha^{ab},} where $[...]$ denotes the
anti-symmetrization over the $a,b$ indices. The factor of
$t^{-1}$ comes from the usual scaling of vielbeins, when we
extract out the radial part $\sqrt{t}$. For the sake of
completeness we refer the reader to \bbdg, where a detailed
discussion of this and other related issues were presented. We
are also assuming, that the traces of the holonomy matrices
$M_{ab}$ are non-zero and real constants. So our analysis herein
will only work, if the above two conditions are met. In fact, the
second condition can be partially relaxed, as we demonstrate
later. Taking all these into account, the cubic term in the
anomaly relation \cubic\ will contribute \eqn\cubeterm{ {\rm
tr}~( {\tilde G}\wedge {\tilde G}\wedge {\tilde G}) = h^3~ {\rm
Tr}(M^{ab} M^{cd} M^{ef}) ~\alpha^{ab} \wedge
\alpha^{cd}\wedge\alpha^{ef}  \equiv h^3 t^{-3},} in accordance
with \trel. The traces of the holonomy matrices are in general
non-zero. The special case, when these become zero was discussed
in \bbdg, where it was shown, that the radial modulus can still
be stabilized and it's value can be explicitly evaluated. Let us
consider a simple toy example, where we
choose the holonomy matrices as $\sigma^{ij}$ with $\sigma^i$
being the Pauli matrices, so that (we take $i,j,k=1, 2, 3$)
\eqn\pauli{\sigma^{12}=\pmatrix{2i&0\cr 0&-2i}, ~~~~\sigma^{23}=
\pmatrix{ 0&2i\cr 2i&0}, ~~~~ \sigma^{31} = \pmatrix{0&2\cr
-2&0}.} We can now calculate the traces by taking into account
the anti-symmetrization of $\alpha^{ij}$. There appear six sums,
all of which are the same. The final result can be written in
terms of $\alpha^{ij}$. For a manifold that is approximately
flat (i.e a manifold, which has an orbifold base) one can show,
that all the $\alpha^{ij}$ are numerically 1. This will give us
the value of the radius of the six-manifold $t_o$ as
\eqn\actrad{t_o = 0.2184~t = 0.2812~(\alpha'|f|^2)^{1/3}.} In
deriving this result we have been a little sloppy. We took the
order $\alpha'^2$ term into account in \potrad. As we know, this
term will receive corrections from $d{\tilde G}$ terms. Let us therefore
tentatively write the additional contributions to the potential
as ${n \alpha'^2 f^6\o t^6}$, where $n$ is an integer. If we
assume $n$ to be small, then the contribution to \radnow\ will be
(in units of $(\alpha'|f|^2)^{1/3}$) \eqn\radaftercorr{t = 1.28763
+0.02565 n - 0.00185 n^2 + 0.000183 n^3 - 2.04 \times 10^{-5}n^4
+ {\cal O}(n^5),} which shows, that the results \radnow\ or
\actrad\ are reliable. Of course, the calculation done in this
simple example is not for the holonomy of the manifolds we are
interested in. So an important question would be whether the
radius can be stabilized to a finite value, after we incorporate
all the additional $\alpha'$ dependences, as well as the
dependence on the holonomy matrices. To answer this question, the
full iterative answer for $G$ has to be calculated. This is
unfortunately very complicated. However, if we do an expansion in
$\alpha'$ (with the choice of roots in \roots), we can show, that
the contribution from higher order terms to \radnow\ slowly
becomes smaller and smaller.

Let us remark, that we can determine the form of the constants
$\gamma$ and $\beta$ appearing in this section. However, in order
to do this, we need the value of the radius $t$. Since the
calculations done above are to leading order in $\alpha'$, we
will assume, that the radius is fixed (to all orders in
$\alpha'$) in terms of the flux density as
\eqn\radfix{t = m ~(\alpha' |f|^2)^{1/3},}
where $m$ is a finite constant. The fact, that this could be
greater than 1, can be seen by incorporating a few higher order
corrections. The values of $\gamma$ and $\beta$, that we get using
the above value of the radius are \eqn\gammabeta{\gamma = -{2\o
c}~{\Big (} m^3 +{3\o 8}{\Big )}, ~~~~~ \beta = {c \o 2 m^{3/2}}.}

Finally, let us remark, that in this section we have made many
simplfying assumptions, in order to determine the value of the
radial modulus. In the next section we will pursue a more detailed
analysis, that can be directly related to the non-K\"ahler
manifolds discussed in the literature and where no simplifying
assumptions will be done.

\newsec{Detailed Analysis of Radial Modulus Stabilization}

In the previous section, even though we have done a precise
calculation, our analysis is still incomplete, because we have
used many simplifying assumptions. In this section we shall
perform an analysis, that is valid for heterotic string
compactifications on non-K\"ahler manifolds, without making any
simplifying assumptions. Some aspects of this have been already
discussed in \bbdg.

To do this calculation, we will take all the components of the
three-form $G$ into account. Since the real part of the
three-form ${\cal H}$ is either a (2,1), (1,2), (3,0) or (0,3)
form, we shall denote the various components as
\eqn\varcomp{G_{\bar 1 2 \bar 3} = h_1, ~~~~G_{1 \bar 2 \bar 3} =
{ h}_2, ~~~~G_{1 2 3}= {h}_3,} and their complex conjugates as
${\bar h}_i, i = 1,2,3$. Supersymmetry requires, that $h_3 =
{\bar h}_3 = 0$, so in the final result these components will
become zero. The reason of why we retain the components with legs
along the $z^3$ and ${\bar z}^3$ directions, is because in the
T-dual Type IIB picture these components are the ones, that
survive the orientifold projection. We will denote analogously
the components of the spin-connection tensor\foot{In this section
the spin-connection is the usual spin-connection $\omega_o$ and
not the effective spin-connection $\omega$.} as $\omega_{oi}$ and
${\bar\omega}_{oi}$ with $i=1,2,3$. Let us now derive the
equation for the three-form $G_{\bar 1 2 \bar 3}$. First, we will
need the one-forms, that can be constructed from the three-forms
appearing in \varcomp, as described in \onefor. They are
explicitly given as \eqn\three{\eqalign{&{\tilde G}^{~ab}_{\bar
1}=2 h_1 e^{2[a}e^{b]\bar 3} + 2 {\bar h}_2  e^{2[a}e^{b] 3} + 2
{\bar h}_3  e^{\bar 2[a}e^{b]\bar 3}, \cr & {\tilde G}^{~cd}_{2}=
2 h_1 e^{\bar 3[c}e^{d]\bar 1} + 2 {\bar h}_2  e^{\bar 3[c}e^{d]
1} + 2 {h}_3 e^{\bar 3[c}e^{d] 3}, \cr & {\tilde G}^{~ef}_{\bar
3}= 2 h_1 e^{\bar 1 [e}e^{f] 2} + 2 {h}_2  e^{1 [e}e^{f] \bar 2}
+ 2 {\bar h}_3 e^{\bar 1[e}e^{f]\bar 2}.}} Similar results can be
written for the one-form spin-connection $\omega_{oi}^{~~ab}$. We
will again be ignoring the $d\omega_o$ and $d{\tilde G}$
contributions, as we will work only to order $\alpha'$. We will
however continue to assume, that $e^{a\mu} = t^{-1/2} e_{\rm o}^{a
\mu}$ and therefore the results will again be valid locally. The
above considerations will tell us, that the equation satisfied by
the component $G_{\bar 1 2 \bar 3}$ takes the form
\eqn\baronetwo{ G_{\bar 1 2 \bar 3} + {\alpha'\o 2} {\rm tr}
{\Big [} -{1\o 3} \omega_o \wedge \tilde G \wedge \tilde G + {2\o
3} \tilde G \wedge \omega_o \wedge \omega_o + {1\o 6} \tilde G
\wedge \tilde G \wedge \tilde G {\Big ]}_{\bar 1 2 \bar 3} =
f_{\bar 1 2 \bar 3},} where $f_{ijk}$ is defined earlier in
\defoff. For the
non-K\"ahler metric studied in \sav,\beckerD,\GP\ and \bbdg, the
base manifold was an orbifold. As a result, the only non-trivial
factors in the metric are the warp factors. This would
mean, that up to powers of the warp factor all components of the
spin-connection are the same\foot{We are ignoring an important
subtlety here. The fiber doesn't scale with the warp factor as we
saw in \nkmet. Therefore, the size of the six-manifold should be
expressed in terms of $r_1$ (the radius of the fiber) and $r_2$
(the radius of the base). For the time being, we shall ignore
this subtlety, as this doesn't affect the final result. We will
consider this towards the end of the paper. Therefore our analysis is
done for the case $\Delta_1 = \Delta_2 = \Delta$ and $\alpha,
\beta$ as constants locally, in \nkmet.}.
Let us therefore
take \eqn\wetook{\omega_{oi}^{~~ab} = f_i(\Delta)
\omega_o \epsilon^{ab},} where $f_i$ needs to be worked out
for every components individually and $\epsilon^{ab}$ is the
antisymmetric tensor. As we discussed in \bbdg, if we consider the set of
equations \three\ to the leading order in $\alpha'$, we can
replace the complex three-forms $h_2, h_3$ by their real parts.
Therefore, up to proportionality constants, we can write the
above three one-forms as $h_1 + \alpha_1, h_1 + \alpha_2, h_1 +
\alpha_3$, where $\alpha_i$ can be easily calculated from the
vielbeins $e^{ai}_{\rm o}$. This will transform the cubic equation
\baronetwo\ into\foot{For more details, see section $4.4 b$ of
\bbdg.} \eqn\icecube{{h_1 t^3\o \alpha'} - {\cal A}_1 \omega_o
(h_1 + \beta_1)(h_1 + \beta_2) + {\cal A}_2 \omega_o^2 (h_1 +
\beta_3) + (h_1 + \beta_4) (h_1 + \beta_5)(h_1 + \beta_6) = {f
t^3 \o \alpha'},} where it is an easy exercise to relate
$\alpha_i$ and $f_i(\Delta)$ to $\beta_i, {\cal A}_1$ and
${\cal A}_2$. We have absorbed the traces of the holonomy
matrices discussed in the previous section into the definition of
$t$. One can also check, that similar equations hold for all
other components of $G$. Therefore, the generic cubic equation,
that we get is \eqn\gencubb{h_1^3 + m h_1^2 + n h_1 + s = 0,}
where $m,n$ and $s$ are integers given in terms of $f, t,
\beta_i$ and ${\cal A}_j$ as \eqn\mensa{\eqalign{& m = \beta_4 +
\beta_5 + \beta_6 -{\cal A}_1 \omega_o, \cr & s = -{ft^3\o
\alpha'}+ \beta_4\beta_5\beta_6 - {\cal A}_1 \omega_o
\beta_1\beta_2 + {\cal A}_2 \omega_o^2 \beta_3 ,\cr & n = {t^3 \o
\alpha'} + \beta_4\beta_5 + \beta_5\beta_6 + \beta_6 \beta_4 -
{\cal A}_1 \omega_o (\beta_1 + \beta_2) + {\cal A}_2
\omega_o^2.}} There is a quadratic term in this equation, that
can be removed by shifting the three-form $h_1$ by $h_1 = h_1 -
{m\o 3}$. This will give us precisely the cubic equation
\cubicnow, with $p$ and $q$ defined as \eqn\pnowq{p = {t^3 \o
\alpha'} - {\cal A}_3 {\hat\omega}_o + {\cal
O}(\hat\omega_o^2),~~~~ q = -{ft^3 \o \alpha'} + {\cal A}_4
{\hat\omega}_o + {\cal A}_5 + {\cal O} (\hat\omega_o^2),} where
$\hat\omega_o$ is a shifted spin-connection, that is introduced
to absorb the constant term in $p$ and ${\cal A}_i$ are constants
determined by \mensa. To order $\hat\omega_o$ the above
expression precisely coincides with the simplified discussion
presented in the previous section (see eq.
\newpandq\ and identify $\hat\omega_o$ with $\omega_o$ there).
There is one difference though, that we would like to discuss in
some detail. Notice, that in the expression for $q$ given above,
there is a constant ${\cal A}_5$, that did not appear in our
analysis in the previous section. This constant is rather
harmless, because adding a constant $l$ into the definition of $q$
in
\newpandq, will give an additional constants in the real and the imaginary
parts of $G$ in \comroot, that have the form
\eqn\realandima{-{l\alpha'\o t^3} + {\cal O}(\alpha'^2), ~~~~{\rm
and}~~~~ 0 + {\cal O}(\alpha'^{3/2}),} respectively. The above
shift appearing in the real root doesn't change the final
expression for $G$, because we can tune $l$ to scale smaller than
${f^3}$. To summarize, what we have just shown is, that we can
trust the analysis done in the previous section, even if some
simplifying assumptions were done.

There is one more point, that we would like to discuss, before we
derive the torsional constraints from our superpotential in the
next section. This has to do with the signs of $p$ and $q$ in the
cubic equation \cubicnow. According to the conventions, that we
have chosen, the signs of $p$ and $q$ are fixed. However, is an
interesting question to ask, what happens, if we reverse the signs
of $p$ and $q$. Observe, that transforming $q \to -q$ does
not change much the real and complex solutions for $G$, as they
still retain their original form, except for an overall sign
change in the function \Aseries. However, changing $p \to -p$
makes \Bseries\ pure imaginary and therefore all the roots of the
cubic equation become real. One may now wonder, if this is
consistent with dual Type IIB picture.

To see, that this is indeed so we need to remind ourselves, that
the tensor field $H_{RR}$ of the Type IIB theory turns into the
real root ${\cal H}$ of the heterotic theory and the tensor field
$H_{NS}$ turns into the spin-connection. Let us therefore make the
transformation \eqn\hrrhns{H_{RR} \to i H_{RR}, ~~~~{\rm and}~~~~
H_{NS}\to i H_{NS},} which shifts the $i$ in the superpotential
to the $H_{RR}$ part, while giving a relative minus sign between
the three-forms. Recall, that we are considering the case, when
we have a vanishing axion-dilaton, as this is directly related to
the superpotential of the heterotic theory. The above
transformation \hrrhns\ will, in turn, convert our cubic equation
\cubicnow\ into \eqn\cunowis{G^3 - p~ G + i q = 0,} where we have
taken $G \to i G$ in \cubicnow. The roots of the above equation
are, in fact, slightly different, because the real root \Aseries\
will become purely imaginary, while the $i$ in the complex root
\comroot\ will trade places. A simple way to see this, would be to
go back to the series expansions \Aseries\ and \Bseries\ and
write them as \eqn\writetem{\eqalign{&A+B = \sum_{n=0}^\infty~a_n
{(-1)^{n+1} |q|^{2n+1}\o p^{3n+1}},\cr &A - B = \sqrt{p} +
\sum_{n=1}^\infty~b_n {(-1)^{n+1} |q|^{2n}\o p^{3n - {1\o 2}}},}}
where $a_n, b_n$ can be determined from \Aseries\ and \Bseries.
Of course, this expansion is not very meaningful beyond the first
few orders of $p$ and $q$, but we shall still use these series, to
illustrate the generic behavior (we believe that putting higher
order $\alpha'$ corrections to the system will change the
coefficients $a_n$ and $b_n$, without altering the $p,q$
behavior). If we change $p \to -p$ and $q \to iq$, it is easy to
check from \writetem, that both $A\pm B$ become purely imaginary,
as mentioned above. Let us now make the transformation $q \to
iq$, which will give us \eqn\qtoiq{G^3 - p~ G - q = 0, ~~~~~ p
> 0, ~~q > 0,} whose roots are all real\foot{The three real roots
are given by $2 a~cos~{\theta \o 3}$ and $-a( cos~{\theta\o 3}
\pm \sqrt{3}~ sin~{\theta\o 3})$, where we have defined $a =
{\sqrt{p\o 3}}$ and $cos~\theta =  {q\o 2}\sqrt{27\o p^3}$.}.
This is the case alluded to above. The question now is to trace
this back to the Type IIB theory.

On the heterotic side, following the expansions \writetem\ and
performing the transformation $q \to iq$, makes the real root $A +
B$ pure imaginary. One should be slightly careful here, because
$q$ defined in the expansions of $A$ and $B$ involve $\sqrt{q^2}$,
which can have any sign. Therefore, ${\cal H}$ becomes purely
imaginary. Since ${\cal H}$ is directly the T-dual of $H_{RR}$,
this would imply, that $H_{RR}$ goes back to the real value. And
therefore, since $H_{NS}$ is pure imaginary, the factor of $i$ in
the three-form $G$ transforms this into a purely real three-form,
exactly as we expected from the heterotic side! In other words, to
make a transition from \cubicnow\ to the cubic equation \qtoiq\
in the heterotic theory, we need to make the transformation
\eqn\iibtrans{H_{RR} \to - H_{RR}, ~~~~ {\rm and} ~~~~ H_{NS} \to
i H_{NS},} in the Type IIB theory for the conventions, that we are
following. This will lead to a real three-form in the Type IIB
theory, exactly as we have in the heterotic theory.

This concludes our discussion regarding the form of the
superpotential involving the three-form flux for compactifications
of the heterotic string on non-K\"ahler complex
six-dimensional manifolds. In the next section we shall see, that
there is one more term in the superpotential, if the effect of the
non-abelian gauge fields is taken into account. Using the complete
superpotential, we shall derive the form of the torsional
constraints.

\newsec{Superpotential and Torsional Constraints}

The goal of this section is to derive the constraints following
from supersymmetry, that compactifications of the heterotic string
on non-K\"ahler complex three-folds have to satisfy. We will do
so, by using the F-terms and D-terms, which describe these sort of
compactifications. Let us first describe the F-terms in full
detail. Until now we have considered the superpotential, that is
written in terms of the three-form $G$. But there is another
superpotential, coming from the heterotic gauge bundle, which we
will compute in this section. This superpotential is distinct
from the Chern-Simons term
\eqn\chercont{  - \alpha^\prime \int \Omega_3({\cal A}) \wedge
\Omega,}
that we have already described and is most easily understood in
terms of the corresponding ${\cal M}$-theory superpotential. In
${\cal M}$-theory on a four-fold ${\cal X}$, $G_4$-fluxes induce
two superpotentials, which can be expressed in terms of the
holomorphic $(4,0)$-form $\Omega_4$ and the K\"ahler form $J$ of
the Calabi-Yau four-fold \guko\
\eqn\mtheorysuper{ W_{\cal M} = \int_{\cal X} G_4 \wedge
\Omega_4,}
and
\eqn\mtheorysupert{ {\hat W}_{\cal M} = \int_{\cal X} J \wedge J
\wedge G_4.}
Demanding $W_{\cal M} = DW_{\cal M} = 0$ and ${\hat W}_{\cal M} =
D{\hat W}_{\cal M} = 0$, reproduces the constraints for unbroken
supersymmetry in Minkowski space derived in \rBB, which state,
that the only non-vanishing component of $G_4$ is the $(2,2)$
component, which has to be primitive.

The first expression \mtheorysuper, is the origin of the
superpotential \superpot, that we have been discussing so far,
because it contributes to the two bulk three-forms $H_{NS}$ and
$H_{RR}$ of the Type IIB theory. Therefore, it contributes to the
heterotic three-form ${\cal H}$ and spin connection $\omega$. The
second expression \mtheorysupert\ gives rise to a superpotential
for the gauge bundle in the Type IIB theory or consequently to a
second superpotential in the heterotic theory. In order to see
this, let us consider ${\cal M}$-theory on $T^4/{\cal I}_4 \times
T^4/{\cal I}_4$, which was discussed earlier in the literature in
\beckerD\ and \bbdg. We decompose the ${\cal M}$-theory flux in a
localized and a non-localized part. The non-localized part is
responsible for the heterotic superpotential \suppot. The
localized part is a little more subtle and it takes the form
\eqn\locform{ {G_4\over 2\pi} = \sum_{i=1}^4~F^i(z^1,z^2,{\bar
z}^1,{\bar z}^2) \wedge \Theta^i(z^3,z^4,{\bar z}^3,{\bar z}^4).}
Here the index `$i$' labels four fixed points and at each fixed
point, there are four singularities. Also, $z^{1,2}$ are the
coordinates of the first $T^4 /{\cal I}_4$ and $z^{3,4}$
correspond to the coordinates of the second $T^4 /{\cal I}_4$
\foot{More details can be extracted from sec 2.5 of \bbdg.}. The
harmonic (1,1)-forms near the fixed points of the orbifold limit
of K3 are denoted by $\Theta^i$. Inserting \locform\ into
\mtheorysupert, we get a contribution to the Type IIB
superpotential, which after integrating out $\Theta$ over a
two-cycle is of the form
\eqn\superadd{\sum_i~\int F^i \wedge J \wedge J.}
The integral over the two-cycle is bounded, since the $\Theta^i$'s
are normalizable. The (1,1)-forms $F^i$ have an interpretation as
gauge fields on the Type IIB side.

Two T-dualities and an S-duality will not modify the above
expression of the superpotential. Therefore, we obtain besides
\suppot\ a second superpotential for the heterotic theory
\eqn\hetsuplook{{\hat W}_{het} = \sum_i \int F^i \wedge J \wedge
J,}
where $J$ is the fundamental (1,1) form of the internal space.
However, this is not the whole story yet. As in the case for
compactifications of the heterotic string on a Calabi-Yau
three-fold, we will not only have F-terms but also D-terms. The
explicit form of these D-terms can be computed from the
supersymmetry transformation of the four-dimensional gluino. This
supersymmetry transformation gives us the following constraints
on the non-abelian two-form of the heterotic theory
\eqn\contso{F^i_{ab} = F^i_{\bar a \bar b} = 0,}
and
\eqn\duy{J^{a\bar b}F^i_{a\bar b}=0,}
as has been explained e.g. in \witsuper. The last equation is the
well known Donaldson-Uhlenbeck-Yau (DUY) equation. These
constraints can be derived from a D-term, appearing in the
four-dimensional theory
\eqn\dterm{D^i=F^i_{mn}J^{mn} = \epsilon^\dagger F^i_{mn} \Gamma^{mn}
\epsilon,}
as supersymmetry demands $D^i= 0 = F^i_{mn}\epsilon^ \dagger
\Gamma^{mn} \epsilon$. However, it turns out, that these
constraints on the gauge bundle can also be derived from the
superpotential \hetsuplook. This is because, it has been shown in
\witsuper, that the following identity holds
\eqn\witeq{F^i_{a{\bar b}}J^{a {\bar b}}={1\over 2} F\wedge
J\wedge J,}
so that the DUY equation is equivalent to the supersymmetry
constraint ${\hat W}_{het}=0$, while $D{\hat W}_{het}=0$ imposes
no additional constraint. From a different perspective notice,
that primitivity of the ${\cal M}$-theory flux translates on the
heterotic side into the previous conditions for the localized
fluxes. As we will see below, the gauge bundle is further
restricted. This additional condition comes from the three-form
part of the superpotential.

We now would like to use the above results to derive the torsional
equations of \rstrom,\HULL\ and \SmitD, which are required for
supersymmetry, from the superpotential
\eqn\superpot{
 W_{het} = \int G \wedge \Omega,}
involving the complex three-form. Notice, that we have related
this superpotential directly to the T-dual of the Type IIB
superpotential. In particular, this implies, that the three-form
$G$ in the heterotic theory should be imaginary self-dual,
because the corresponding T-dual configuration in IIB is! This
means, that one condition for unbroken supersymmetry for
compactifications of the heterotic string to four-dimensional
Minkowski space is
\eqn\seldd{ \star_6({\cal H} + i\beta \omega) = i ({\cal H} + i
\beta \omega),}
where the Hodge $\star_6$-operator is defined with respect to the
six-dimensional internal manifold. Observe, that this is the gauge
invariant three-form, which satisfies the Bianchi identity with
respect to the connection with torsion.

Comparing the real and the imaginary sides of the above equation,
we obtain the condition
\eqn\toreqn{ {\cal H} = \pm \star_6 \beta ~ \omega,}
where we have kept the sign ambiguity, to reflect the fact, that
the real three-form can have either sign in this space. This is
basically the content of the torsional equation, which we shall
write now in the more familiar form appearing in \rstrom,\HULL\
and \SmitD. Our goal is to express this constraint in terms of the
fundamental two form $J_{mn}$, where $m,n$ are spatial
coordinates on the non-K\"ahler space. But before we do that, we
need to carefully define the spin-connection $\omega$. Recall,
that this spin connection is the {\it effective} spin-connection
given in \omegeff\ and therefore we have to be slightly careful in
defining it. From the form of \omegeff\ we can see, that it has a
piece proportional to $\omega_o$ and a piece proportional to
$|f|$ plus higher order corrections. Furthermore, being a
three-form it is completely antisymmetric in all of its three
space-time indices. In terms of the vielbein therefore, it should
be an anti-symmetric combination of $e$ and  $\del e$ for
dimensional reasons. Let us therefore write the complete
antisymmetric part of $\omega$ as\foot{Recall that this effective
spin-connection can have contributions from purely covariant terms. The 
result below is when we take into account all the possible 
geometric and non-geometric terms.}
\eqn\spincon{\omega_{[nml]} =
G_1~ \eta_{ab}~ e^a_{[n}\del_{m}e_{l]}^b + G_2~ \epsilon_{ab} ~
e^a_{[n}\del_{m}e_{l]}^b,} where $a,b$ are internal indices and
$G_{1,2}$ are, in general, functions of the warp factors. We
should also view this form of the spin-connection only locally,
as the vielbeins are defined locally, because there are no
one-forms on our non-K\"ahler spaces.
 There would be fermion contributions to the
above formula (as given in eq 17.12 of \weba), but we are ignoring
them for the time being. The above relation will imply, that we
can write \toreqn\ equivalently as \eqn\toernow{{\cal H}_{mnp} =
\beta~\sqrt{g}~
 \epsilon_{mnp}^{~~~~~~qrs} \omega_{qrs}.}
At this point we can use the specific background, that we have
been taking all through, which is given by \cubicnow\ and
\assumpta\ to determine the possible values of $G_{1,2}$. First,
it is easy to check, that the $G_1$ dependent term vanishes for
this choice of background. Notice again, that our analysis is
valid only, if the metric is of this form locally\foot{It is an
interesting question to obtain the full global picture, by taking
$\alpha$ and $\beta$ to be non constant in \nkmet. In this case
the simple analysis of the cubic equation will no longer be valid
and we have to do a more precise evaluation. The ansatz for the
spin-connection made in \wetook\ will no longer be valid either.
This will affect the complex part and the real part of the
three-form \comroot. We hope to address this elsewhere
\toappear.}. In the presence of warp factors we would in principle
expect the definition of the vielbeins to get modified. But, as
we will soon see, the form of the vielbeins still remains as
above, but now with a modified $t$.

Therefore, in the absence of warp factors we are left with the
second term in \spincon\ with a constant $G_2$. Introducing back
the warp factors will not change this conclusion. Now equation
\toernow\ has almost the form, in which the torsional equation
appears in \rstrom, \HULL\ and \SmitD\ (see e.g. equation 3.49 of
\SmitD), but it is still formulated in terms of the
spin-connection instead of the fundamental two-form. We are
ignoring the warp factors, so we do not see the dilaton explicitly
in the formula.

Let us rewrite \toernow\ in terms of the fundamental two-form $J$.
This is not difficult, as the term proportional to $G_2$ is simply
$dJ$, where $J$ is the usual two-form of the manifold. The precise
relation between $G_2$ and $dJ$ can be derived in the following
way. Let us first define a covariantly constant orthogonal matrix
${\rm N}$, such that ${\rm N}^\top = {\rm N}^{-1}$ and this would
convert the $D_4$ spinor indices (world-sheet indices) to vector
indices. More details on the sigma model description of the
heterotic string on non-K\"ahler manifolds have appeared in
section 2.4 of \bbdg. We will follow the notations of that
section. This means \eqn\means{S^a = {\rm N}^a_{~~q} S^q,
~~~~~S^i= e^i_{~~a}S^a,} where $S^p, p = 1, ...,8$ is the
world-sheet superpartner of $X^i$, describing the light cone
coordinates\foot{Recall, that we are imposing $S^{\dot q} = 0$,
therefore only 8 components remain. The gamma matrix $\Gamma^i_{p
\dot q}$ acts as triality coefficients, that relate the three
inequivalent representaions of $D_4$, i.e the vector and the two
spinor representations.}. Therefore ${\rm N}$ is an $8 \times 8$
matrix described in more detail in \greenS\ and \HULL. Following
earlier work, we can choose the ${\rm N}$ matrices as
antisymmetric, such that the two-form is given by \HULL
\eqn\twfor{J_{ij} = {\rm N}_{ab} ~e^a_{[i}e^b_{j]}.} Now we can
use the epsilon tensor of the Hodge $\star$ to rewrite the right
hand side of \toernow, involving the spin-connection
$\omega_{qrs}$ in terms of derivatives acting on the vielbeins as
\eqn\spinnow{{\cal H}_{mnp} =
\sqrt{g}~{\beta}~G_2~\epsilon_{mnp}^{~~~~~qrs} ~ \epsilon_{ab}~
e^a_{q} \del_r e^b_s,} where we have ignored a factor of 6, as we
are more interested in the functional dependences. In deriving
this, we have used the explicit form of the spin-connection given
in \spincon. Now equation \spinnow\ suggests, that the right hand
side can be expressed in terms of the derivative of the two-form
$J_{ij}$ appearing in \twfor. In fact, we can exploit the
antisymmetry of ${N}$ to express this as \eqn\jform{[\star
dJ]_{mnp} = \sqrt{g}~\epsilon_{mnp}^{~~~~~qrs}~ {\rm
N}_{ab}~\del_q (e^a_{[r}e^b_{s]}) = \sqrt{g}~
\epsilon_{mnp}^{~~~~~qrs}~ N_{ab}~ e^a_{q}\del_r e^b_s,} where
there would again be a proportionality constant, that we are
ignoring. From \spinnow\ and \jform, we require
$\beta~G_2~\epsilon_{ab} = 2 N_{ab}$. The constant $G_2$ can then
be easily worked out from the known expressions of ${ N}$ and
$\beta$. Notice, that in the calculation above we have always been
taking simple derivatives, while we should have taken covariant
derivatives. The connection appearing in this covariant
derivative should be the Christoffel connection and not the
torsional connection. More details of this calculation are given
in \SmitD, so we refer the reader to this paper for further
information. After taking this into account, the torsional
equation, that we obtain from our superpotential \superpot\ is
\eqn\torderived{{\cal H}_{mnp} =
\sqrt{g}~\epsilon_{mnpqrs}~D^qJ^{rs},} where $D_p$ is the
covariant derivative. This is consistent with the result of the
earlier literature \rstrom,\HULL\ and \SmitD, when we have no warp
factor. What happens, when we introduce back the warp factor? To
see this recall, that the warp factor in the heterotic theory is
proportional to coupling constant, i.e $\Delta = e^\phi$, with
$\phi$ being the heterotic dilaton. In fact, both ${\cal H}$  and
$J$ will scale in some particular way with the warp factor giving
us the following dilaton dependence of the torsional equation:
\eqn\wardil{{\cal H}_{mnp} = \sqrt{g}~e^{a\phi}
\epsilon_{mnpqrs}~D^q (e^{b\phi} J^{rs}),} where $a$ and $b$ can
be determined for our case when $\Delta_1 = \Delta_2 = \Delta$.
This gives us precisely the result, we had been looking for. This
equation has recently also been discussed in \gauntlett, where
the values of $a,b$ in \wardil\ were derived from the Killing
spinor equations and not using any superpotential. For our case,
when we restrict ourselves to the more realistic scenario of
\nkmet, we can view the radial modulus $t \equiv t(x,y)$ with
$x$, the coordinates of 4d Minkowski space-time and $y$, the
coordinates of the internal non-K\"ahler space, as
\eqn\actt{t(x,y) = {\tilde t}(x)~\Delta^2(y)} so that the warp
factor is {\it absorbed} in the definition of $t$ itself and
${\tilde t}$ is the ``usual'' radial modulus that is independent
of the coordinates of the internal manifold. This way of looking
at things tells us the reason why in the presence of warp factors
(at least for the conformal case) we expect our analysis to go
through. In fact for the case where we have $\Delta_1 = \Delta,
\Delta_2 = 1$, the
 above choice of $t$ will tell us that we can take (instead of
$\Delta_1 = \Delta_2 =1$) a slightly more involved case where
$\Delta_1 = 1, \Delta_2 = \Delta^{-2}$. This analysis, now with
$\alpha, \beta$ no longer constant in \nkmet, will be dealt in
\toappear. However various indirect arguments suggest that the
values of $a,b$ in \wardil\ are given by $a = b = 2$. Therefore we
can now write \wardil\ in a condensed way as \eqn\wadi{{\cal H} =
e^{2\phi} \star_6 d(e^{-2\phi} J).} One can show, that this form
of the torsional equation is identical to the conventional form
\eqn\torconv{{\cal H} = i (\del - \bar\del)J,} when specified to
the case of compactifications with $SU(3)$ structure. To show
this, we will use the description of manifolds with $SU(3)$
structure suggested in \GP.\foot{The analysis of \gauntlett\
which uses Killing spinor equations {\it a- la} ref. \rstrom\
does not require any specific background. In fact even though we
use some background to illustrate the torsional equation and the
radial stabilisation, our analysis is completely general as the
technique of cubic equation discussed above {\it does not}
require any specific input. Furthermore, though not directly
related to our interest here, many new examples of non-compact
geometries are studied in \gauntlett\ preserving different
fractions of susy (see the third reference of \gauntlett). We
thank the authors of \gauntlett\ for informing us about these
interesting developments.} In the notations of \GP, the metric
\nkmet, \earlywor\ has the form \eqn\gpmetric{g = e^{2\phi}
\pi^\ast g_{CY} + \rho \otimes {\bar\rho},} where $g_{CY}$ is the
metric of the Calabi-Yau base, in our case it is K3 or its
orbifold limit $T^4/{\cal I}_4$. Also, $\phi$ is a function on
the base Calabi-Yau because recall, that it is related to the
warp factor as $\Delta = e^\phi$ and $\Delta$ is a function of
the base, for the examples studied in \sav,\beckerD\ and \bbdg.
Finally, the (1,0) form $\rho$ is such that \eqn\defofrho{d\rho
\equiv \sigma = \omega_P + i \omega_Q,} with the real (1,1) forms
on the Calabi-Yau base defined as \eqn\realformoncy{\eqalign{&
\omega_P = 2~d {\bar z}^2 \wedge dz^1 - 2~ d {\bar z}^1 \wedge
dz^2, \cr & \omega_Q = (1+i) ~ d {\bar z}^2 \wedge dz^1 - (1-i)~d
{\bar z}^1 \wedge dz^2,}} obeying the susy condition\foot{We take
$\star$ to be an anti-linear operation.}: $\star_4 \omega_{P,Q} =
- \omega_{P,Q}$. These (1,1) forms are basically related to the
fibered metric defined in \nkmet, as one can easily extract, by
comparing \gpmetric\ and \nkmet. For the fundamental form we have
\eqn\funform{\eqalign{&J = e^{2\phi} \pi^\ast J_{CY} + {i\o 2}
\rho \wedge {\bar \rho},\cr & dJ = 2 e^{2\phi} d\phi
\wedge\pi^\ast J_{CY} + {i\o 2} (\sigma \wedge \bar\rho -
\bar\sigma \wedge \rho),}} where we have used $dJ_{CY} = 0$. Now
we can use the special properties of $\sigma$ and $\rho$ with
respect to the Hodge star operations \eqn\hodgee{\star_4 \sigma =
- \sigma, \qquad \star_2 \rho = i \bar\rho, \qquad \star_2
\bar\rho = - i \rho,} to get the following set of relations
\eqn\setofrel{\eqalign{& (1)~~ i(\del - \bar\del)J = i
\left[(dJ)^{2,1} - (dJ)^{1,2} \right] = i(\del -
\bar\del)e^{2\phi} \wedge \pi^\ast J_{CY} + \star_6 {i\o 2}
d(\rho \wedge \bar\rho), \cr & (2)~~ e^{2\phi}\star_6
d(e^{-2\phi}J) = - \star_6 \left( 2 d\phi \wedge {i\o 2}\rho
\wedge \bar\rho \right) + \star_6 d \left( {i\o 2}\rho \wedge
\bar\rho \right).}} We can now use the distributive properties of
the Hodge star, to write the first term in the second relation of
\setofrel\ as \eqn\dishod{\star_4 (2~ d\phi) \wedge \star_2
\left({i\o 2} \rho \wedge \bar\rho \right) = \star_4 ( 2~ d\phi)
= i (\del - \bar\del) e^{\2\phi} \wedge \pi^\ast J_{CY}.}
Plugging \dishod\ into \setofrel, we can easily see that
\eqn\easytosee{ i(\del - \bar\del)J = e^{2\phi} \star_6
d(e^{-2\phi} J),} thus proving the relation \torconv. This
relation is, of course, reflection of the fact, that the two-form
$J$ is ${\cal H}$-covariantly constant with respect to the
modified connection, which includes the torsion. There is a
factor of ${1\o 2}$ in \torconv\ different from the result
presented in \rstrom. This has already been shown in \bbdg\ to be
a consequence of the choice of the real three-form in the
connection as ${1\o 2} {\cal H}$, instead of just ${\cal H}$. It
is also clear from the relation \wadi, that $dJ \ne 0$, so that
the six-manifold is non-K\"ahler. This property is directly
related to the fact, that the fibration given by the metric
\gpmetric\ is non-trivial. The (1,0) form $\rho$, corresponding
to the $T^2$-fiber, is not (globally) closed, and there is a
mixture between the fiber and the base coordinates. As a
consequence, the right hand side in the second formula in
\funform\ is non-zero and $dJ \ne 0$.

Another way to see this relation, is to use the fact, that the
holomorphic $T^2$-fiber is a torsional cycle and is zero in the
real homology of the six manifold \GP\ and \bbdg. It is similar
to the one-cycle in the $\IR\IP^2$ example pictured in fig. 4.
Indeed, suppose $dJ = 0$. The integral $\int_{T^2} J$ over the
fiber $T^2$ gives its volume and must be non-zero. On the other
hand, an integral of any closed form over a torsional cycle is
zero. Therefore, we conclude $dJ \ne 0$.

\bigskip

\centerline{\epsfbox{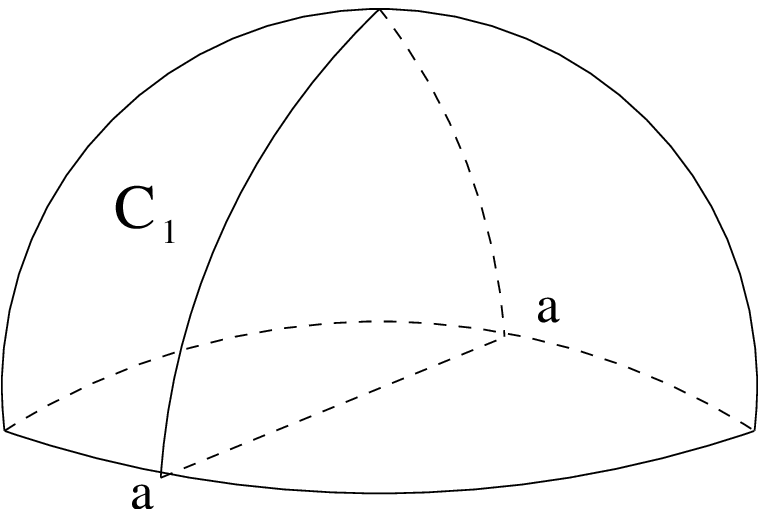}}\nobreak \centerline{{\bf Fig.
4}: {\it The torsional cycle $C_1$ on $\IR\IP^2$ is $\IZ_2$ in the integer
homology}} \centerline{{\it and zero in the real homology.}}

\bigskip

The above result \wardil\ is in string frame. In Einstein frame
we expect $N^{ab}$ to change in the following way \eqn\eien{{\rm
N}^{ab}~ \to ~ e^{g\phi}~{\rm N}^{ab},} where $g$ is a constant
and therefore \jform\ will pick up an additive piece proportional
to $\star \del^q \phi J^{rs}$ from the fact, that the metric in
string frame is related to the metric in Einstein frame via the
relation $g^{\rm string}_{\mu\nu} = e^{\phi\o 2} g^{\rm
Einstein}_{\mu\nu}$. This would imply, that the torsional equation
in Einstein frame becomes \eqn\torein{{\cal H}_{mnp} = e^{h\phi}
\sqrt{g}~\epsilon_{mnpqrs}[ D^q J^{rs} + c~\del^q \phi J^{rs}],}
where $h$ and $c$ are constants, that can be easily determined by
carefully studying the transformation rules from the string frame
to the Einstein frame. This is precisely the form, in which the
torsional equations appear in \SmitD. Observe, that the
superpotential analysis gave a very simple derivation of this
relation. Furthermore, on our six-dimensional space the two form
$J$ satisfies: $\star J = {1\o 2} J \wedge J$. This implies, that
the Nijenhuis tensor vanishes \SmitD.

{}From the above torsional equation it is now easy to extract the
additional constraint on the gauge bundle. We have already shown,
that the gauge bundle in this space has to satisfy the
Donaldson-Uhlenbeck-Yau equation. The torsional equations derived
above show, that there is a further constraint (alluded to earlier
in \bbdg) given by \eqn\duycons{{\rm Tr}~F \wedge F = {\rm tr}~R
\wedge R - i \del {\bar\del} J,} in addition to the ones
presented in \contso\ and \duy. The solutions of these equations
and their phenomenological aspects will be discussed in a
forthcoming paper \toappear. Before we end this section, we make
the following observations.

{}First, from the form of the torsional equation it is  clear,
that we cannot scale the fundamental two-form $J_{mn}$ in an
arbitrary way. This is, of course, related to the stabilization
of the radial modulus for this manifold. The point to note is,
that the torsional equation (which is basically the statement,
that the two-form $J$ is covariantly constant with respect to some
torsional connection) implies, that the modified connection is
contained in $SU(3)$, because the first Chern class vanishes. As
we discussed earlier, this modified connection appears, when we
choose our contorsion tensor to be precisely the torsion (see the
discussion section of \bbdg).

The second observation is related to the sizes of the base K3 and
the fiber torus $T^2$ for our non-K\"ahler manifold\foot{The
following discussion arose from a conversation with
M.~M.~Sheikh-Jabbari. We thank him for many helpful comments.}. In
\radnow\ we fixed the overall radius $t$ of our manifold. If we
call the radius of the fiber $T^2$ as $r_1$ and the radius of the
base K3 as $r_2$, then we have the identity $t^3 = r_1^2 r_2^4$.
Is it now possible to fix both $r_1$ and $r_2$, knowing $t$? In
principle from the choice of the potential \potrad\ we {\it
cannot} fix both. But we can use T-duality arguments to fix the
radius of the fiber. Indeed in \beckerD\ it was shown, that fixing
the Type IIB coupling constant actually fixes the volume of the
fiber $T^2$ to $\alpha'$ (see eq. 3.3 of \beckerD). This would
imply, that for our case we have \eqn\radoffibbase{r_1 =
\sqrt{\alpha'}, \qquad r_2 = \sqrt{|f|},} for the radii of the
fiber and the base respectively. The fact, that the fiber is
stabilized at the value $\alpha'$ is not too surprising, because
we have shown, that our model can be understood from T-duality
rules. Since T-dual models have a self-dual radius at value
$\alpha'$, we expect the same for our case. An important question,
however is now, whether we can trust the supergravity analysis. In
fact, for our case we can only provide an {\it effective}
four-dimensional supergravity description
as long as the {\it total} six-dimensional volume is a
large quantity because of the inherent topology change. 
Since $|f|$, the flux density, can in principle be
large, even though the total flux over a three-cycle ${\cal C}^3$,
which is $\int_{{\cal C}^3} |f|$ is a fixed quantity, we can have
a large sized six-manifold. Observe that the duality chasing arguments 
that we followed to derive the background doesn't rely on the existence of
a supergravity description because the type IIB solution that we took is an 
exact F-theory background. Unfortunately a direct confirmation of this 
is not possible in the heterotic theory because it will 
in principle be difficult to explain the topology change via supergravity
analysis\foot{We thank G. Cardoso, G. Curio, G. Dall'Agata and D. Luest for
correspondence on this issue. See also \ccdl.}.

The third observation is related to the potential $V(t,f)$ for the
radial modulus $t$. One can easily verify, that $V(t,f)$ has the
form of the potential for a half harmonic oscillator, whose minima we
computed in the previous section. This would imply, that the
corresponding Schr\"odinger equation will only have
wave-functions, that vanish around the minima of the potential.
This in turn will determine the spectrum of radial fluctuations
of our system. We can also use \potrad\ to calculate the possible
mass of the radion\foot{We thank S. Kachru for discussions on
this aspect.}. This is basically given by the usual formula:
${\del^2 V\o \del t^2}$, which can be explicitly determined for
our case. In principle this can be a large quantity, because of
the arguments given above.


\newsec{Discussions}
In this paper we have shown, that the perturbative superpotential
for the heterotic string theory compactified on a non-K\"ahler
complex threefold computed in \pktsptt\ and \bbdg, contains an
additional term, if the effect of non-abelian gauge fields is
taken into account. To the order that we took in the anomaly equation,
our proposal for the complete perturbative
superpotential for
heterotic theory compactified on non-K\"ahler manifold ${\cal M}_6$,
when we consider
non-trivial warping and also possible geometric terms in the complex
three-form $G$, is given by (ignoring the warp factors)
\eqn\comsuperpotg{W = \int_{{\cal M}_6}
\left[{\cal H} + i~ dJ \right] \wedge \Omega +
\int_{{\cal M}_6} F \wedge J \wedge J.}
This form of superpotential\foot{The above form of the superpotential (without
the gauge field contribution) has also been derived recently using an
alternative method in an interesting paper \ccdl.}
presumably survives to {\it all} orders in
$\alpha'$ when the full iterative solution is found. Here we give a brief
sketch of the situation when we incorporate all the possible effects.
It is easy to show that the generic form of the
three-form will now be given by the expression
\eqn\genG{G = (a~{\cal H} + \star_6 A) + i~(dJ + B),}
where $A$ and $B$ are generic
functions of $\omega_o$, the torsion-free spin-connection, and $f$ defined
earlier.
In this form we expect $G$ to be anomaly free and gauge invariant, therefore,
upto possible gauge invariant
 terms, this should solve the anomaly equation. For the
case discussed above, i.e. when we considered to the first order in $\alpha'$
 in the anomaly equation, we had
a cubic equation where it was shown explicitly that $A = B = 0$
and therefore $G$ was given as
\eqn\gcubic{G = a~{\cal H} + i~dJ,}
ignoring the warp factors. Now in the presence of $A$ and $B$ the
ISD condition on our background will imply (ignoring possible constants)
\eqn\isd{{\cal H} =  \star_6~dJ + \star_6~(B - A).}
There could be two possibilities now: (a) The torsional constraint derived
earlier and mentioned above is {\it invariant} to all orders in $\alpha'$. In
that case we expect $A =  B$; or (b) The torsional constraint receives
correction. In that case the corrections would be proportional to
$\star_6~(B - A)$. Now, there are ample evidences that suggest that the former
is true and therefore $A$ should be equal to $B$ upto possible gauge invariant
terms. Furthermore if we also demand that
\eqn\condAB{A = B = i \star_6 A}
then both the torsional equation and the superpotential \comsuperpotg\
will be exact to {\it all}
 orders in $\alpha'$. More details on this will appear
in \toappear.

The superpotential \comsuperpotg\ is a
generalization of the superpotential for the heterotic string
theory compactified on a Calabi-Yau threefold, first described in
\witsuper\ and recently in more detail in \bg\ and \becons. An
important consequence is, that due to the presence of this
potential all the complex structure moduli and some K\"ahler
structure moduli get frozen\foot{Alternative framework for freezing some of
the moduli using asymmetric orientifolds or duality
twists have been discussed in \eva. It will be interesting to find the
connection between flux-induced stabilization and these techniques.}.
Furthermore, we have shown, that the
torsional constraint ${\cal H} = i (\del - \bar\del)J$ first
found in \rstrom,\HULL\ and \SmitD\ can be obtained from this
superpotential, in a similar way, as the superpotential of \guko\
reproduces the supersymmetry constraints derived in \rBB. This
torsional constraint implies, that the overall size of the
internal manifold is fixed. Indeed, the previous formula is not
invariant under a rescaling of the K\"ahler form $J$, as the left
hand side is non-zero and frozen to a specific value. A direct
computation of the scalar potential for the radial modulus shows,
that this potential does have a minimum. We have given an
estimate for the value of the radius in terms of the density of
the ${\cal H}$ flux.

There are many interesting directions for future research. Let us
just mention a few. Notice, that the non-K\"ahler manifolds
discussed in this paper all have a vanishing Euler
characteristics. It will be interesting to construct non-K\"ahler
complex manifolds with non-zero Euler characteristics. For this
generalization, we should start with a manifold, which looks like
$K3 \times Z$, where $Z$ is a two-dimensional manifold with
non-zero Euler characteristics on the Type IIB side. This is the
minimal requirement. Of course, we can even get a generic
six-dimensional manifold $X$, which should then have the
following properties in the {\it absence} of fluxes: (a) compact
and complex with non-zero Euler characteristics, (b) there exists
a four-fold, which is a non-trivial $T^2$ fibration over $X$ and
most importantly (c) should have an orientifold setup in the Type
IIB framework. More details on this will be addressed in a future
publication \toappear.

Another direction for research in the future is the following. It
has been shown some time ago in \witsuper, that there are no
perturbative corrections to the superpotential for
compactifications of the heterotic string to four dimensions, but
nevertheless there can be non-perturbative corrections. For the
compactifications considered herein, there are non-perturbative
corrections coming from the dilaton and it would be interesting
to compute their explicit form. More concretely, the
non-perturbative effect, that is directly responsible for the
case at hand is gaugino condensation, which has been studied in
\DineRZ,\rohmwit\ and \maeda\ for compactifications of the
heterotic string on a Calabi-Yau threefold \foot{A more detailed
study of the potential for the heterotic string compactified on a
Calabi-Yau three-fold, taking into account gaugino condensation
will appear in \kg. We thank the authors of this paper for
informing us about their results prior to publication.}. In fact,
the key observation has already been made in the corresponding
Type IIB framework in the presence of fluxes in \renshamit, where
the form of the superpotential in the presence of
non-perturbative effects has been presented. We expect a similar
picture emerges in the heterotic theory compactified on
non-K\"ahler complex manifolds. A detailed discussion will be
presented in \toappear, so we will be brief here.

As observed in \DineRZ\ for ordinary Calabi-Yau compactifications,
the gluino bilinear term ${\rm tr}~{\bar\chi} \Gamma_{\mu\nu\rho}
\chi$ appears in the lagrangian together with the three-from
$H_{\mu\nu\rho}$ as a perfect square. If we denote the gluino
condensate as $\kappa_{\mu\nu\rho}$, then the non-perturbative
contribution to the superpotential is expected to be \eqn\npcontr{
W \sim \int \kappa~ e^{i\alpha + \beta f(\phi)} \wedge \Omega,}
 where $\beta$ may depend on other fields but not on the
dilaton $\phi$ and $f(\phi)$ is some exponential function of
$\phi$. As in \DineRZ\ we have kept a phase $e^{i \alpha}$ (see
\DineRZ\ for more details on this). The above potential will
break supersymmetry,  because the gluino condensate does. Some
details of this analysis have been discussed in \DineRZ\ and
\maeda. In particular, it was shown, that some combination of the
ten-dimensional dilaton and the radial modulus is fixed by this
potential. It will be interesting to apply this mechanism to the
examples studied in this paper, which have a fixed radius of
compactification in order to obtain a model with both the radius
and the dilaton fixed in terms of the expectation values of the
${\cal H}$ flux and the gluino condensate ${\bar\chi}
\Gamma_{\mu\nu\rho} \chi$. This model with stabilized radius and
dilaton could be very useful to study cosmological scenarios,
especially inflation. The T-dual version of this model (in the
Type IIB theory) has been shown to give an interesting
inflationary model \carlos,\renata,\renshamit\foot{We have been
informed that some related work on inflationary scenario which
extends the work of \renshamit\ and \renata\ is currently being
pursued \trivsham.}. It
is plausible, that we can use the same setup, but now for the
non-K\"ahler manifolds considered herein, to construct a
cosmological scenario. We would then have constructed a {\it
rigid} model, that closely simulates some realistic phenomena of
nature. If realized, this would be a major achievement.

\noindent {\bf Note Added}: Recently there appeared an interesting paper which 
discusses the origin of the superpotential \comsuperpotg\ from an alternative
point of view. This paper also discusses the possibility of non-existence
of a supergravity description directly in the heterotic theory, which we 
perfectly agree because of the inherent topology change. We also corrected 
some errorneous statements regarding supergravity description.

\centerline{{\bf Acknowledgements}}

\noindent Its a pleasure to thank Edward Goldstein, Paul Green,
Chris Hull, Shamit Kachru, Renata Kallosh, Andre Linde, Xiao Liu,
Liam McAllister, Ashoke Sen, M.~M. Sheikh-Jabbari and Edward Witten for many
helpful conversations. The work of K.B. is supported by the
University of Utah. The work of M.B. is supported in part by NSF
grant PHY-01-5-23911 and an Alfred Sloan Fellowship. The work of
K.D. is supported in part by a David and Lucile Packard Foundation
Fellowship 2000-13856. The work of S.P. is supported by Stanford
Graduate Fellowship.

\vfill

\break

\vfill

\eject

\listrefs

\bye